\documentclass[a4paper,12pt]{article}

\usepackage{amsthm,amsmath,amssymb,amsfonts,pstricks,pst-node,graphics}

\providecommand*{\pderiv}[3][]{%
        \frac{\partial^{#1}{#2}}%
                {\partial {#3}^{#1}}}

\def\ddt{\pderiv{}{t}}
\def\im{\operatorname{Im}}

\newcommand\A{{\mathcal A}}

\newcommand\la{{\lambda}}
\newcommand\dl{\Delta}

\renewcommand\d{{\rm d}}
\newcommand\D{{\mathcal D}}

\newcommand\J{{\mathcal I}}
\newcommand\cH{{\mathcal H}}

\def\bbbn{{\mathbb N}}
\def\bbbc{{\mathbb C}}
\def\bbbz{{\mathbb Z}}

\newcommand{\lieh}{{\mathfrak{h}}}
\def\f{{\bf f}}
\def\g{{\bf g}}
\def\tr{\mbox{tr}}
\def \rank {\, \mbox{rank}\, }

\newtheorem{Rem}{Remark}
\newtheorem{Not}{Notation}
\newtheorem{Def}{Definition}
\newtheorem{The}{Theorem}
\newtheorem{Pro}{Proposition}
\newtheorem{Lem}{Lemma}
\newtheorem{Ex}{Example}

\begin{document}
\bibliographystyle{unsrt}
\title{Lenard scheme for two dimensional periodic Volterra chain}
\author{Jing Ping Wang\\
Institute of Mathematics and Statistics, University of Kent,\\
Canterbury, Kent, CT2 7NF, United Kingdom}
\date{}  \maketitle

\begin{abstract}
We prove that for compatible weakly nonlocal Hamiltonian and symplectic operators,
hierarchies of infinitely many commuting local symmetries and conservation laws can
be generated under some easily verified conditions no matter whether the generating
Nijenhuis operators are weakly nonlocal or not.
We construct a recursion operator of
the two dimensional periodic Volterra chain from its Lax representation and prove
that it is a Nijenhuis operator. Furthermore we show this system is a (generalised)
bi-Hamiltonian system. Rather surprisingly, the product of its weakly nonlocal
Hamiltonian and symplectic operators gives rise to the square of the recursion
operator.
\end{abstract}
\newpage

\section{Introduction}
Integrable nonlinear evolution equations possess many hidden properties such as
infinitely many symmetries and conservation laws. These symmetries can be generated
by so--called recursion operators \cite{AKNS74,Olv77}, which map a symmetry to a new
symmetry. All known recursion operators including operands \cite{mr94h:35241} for
nonlinear integrable equations in the $2+1$-dimension are Nijenhuis operators.  The
important property of such an operator is to construct an abelian Lie algebra. This
property was independently studied by Fuchssteiner \cite{Fuc79} and Magri
\cite{Mag80}, where they named the operator hereditary symmetry. For example, the
famous Korteweg-de Vries (KdV) equation
$$ u_t=u_{xxx}+6 u u_x$$ possesses a recursion operator
$$\Re=D_x^2+4 u +2 u_x D_x^{-1},
$$
where \(D_x^{-1} \) stands for the left inverse of \(D_x\). Thus this recursion
operator is only defined on \(\im D_x\). It is a Nijenhuis operator and generates the
KdV hierarchy
\[
u_{t_j}=\Re^j (u_x),\qquad j=0, 1, 2,\cdots .
\]
Any polynomial of $\Re$ with constant coefficients such as $\Re^2$ is also a
recursion operator of the KdV. However, the operator $\Re^2$ doesn't generate the
whole KdV hierarchy starting form $u_x$.

The concept of Hamiltonian pairs was introduced by Magri \cite{Mag78}. He found that
some systems admitted two distinct but compatible Hamiltonian structures (Hamiltonian
pairs) and named them twofold Hamiltonian system, nowadays known as bi-Hamiltonian
systems. The KdV equation is a bi-Hamiltonian system. It can be written
\[
u_t=D_x\ \delta (-\frac{u_x^2}{2}+u^3)=(D_x^3+4 u D_x +2 u_x)\ \delta
(\frac{u^2}{2}),
\]
where $\delta$ is variational derivative with respect to the dependent variable.
These two differential operators $D_x$ and $D_x^3+4 u D_x +2 u_x$ form a Hamiltonian
pair.

Interrelations between Hamiltonian pairs and Nijenhuis operators were discovered by
Gel'fand \& Dorfman \cite{GD79} and Fuchssteiner \& Fokas
\cite{mr82g:58039,mr84j:58046}. For example, the Nijenhuis recursion operator of the
KdV equation can be obtained via the Hamiltonian pair, that is,
\[\Re=(D_x^3+4 u D_x +2 u_x)\ D_x^{-1}\ .\]
Such a decomposition of the operator $\Re$ corresponds to the Lenard scheme used to
construct the hierarchies of infinitely many symmetries and cosymmetries.
The story why this concept was named after Lenard is told in \cite{ps05}.

In fact, the decomposition of $\Re$ is not unique since it can also be represented as
$$\Re=D_x\ (D_x+2 u D_x^{-1}+2 D_x^{-1} u),
$$
where the operator $D_x$ is Hamiltonian and the operator $D_x+2 u D_x^{-1}+2 D_x^{-1}
u$ is symplectic.

The majority  of $1+1$-dimensional Hamiltonian integrable equations possess the same
property as the KdV equation: their Nijenhuis recursion operators can be decomposed
into the products of weakly nonlocal \cite{MaN01} Hamiltonian and symplectic
operators of order not less than $-1$ ( see \cite{wang1} for a list of integrable
systems). An exceptional recursion operator can be found in \cite{KS4} although it
can be represented as a ratio of two compatible weakly nonlocal Hamiltonian operators
\cite{serg5b}.

In applications, the Lenard scheme for both Hamiltonian and symplectic pairs
\cite{mr94j:58081} requires that one of the operators is invertible, which is not
clearly defined for (pseudo-)differential operators in the sense that the inverses of
many differential operators are no longer local. For compatible Hamiltonian and
symplectic operators, there is no need to invert any operator in construction of
Nijenhuis operators although the operators considered are likely to be nonlocal. The
nonlocality has motivated Dorfman to introduce Dirac structures to the field of
soliton theory \cite{dorf87}. Dirac structures unify and extend both Hamiltonian and
symplectic structures. She showed that pairs of Dirac structures give rise to
Nijenhuis relations, which is a generalization of the Nijenhuis operators associated
with pairs of Hamiltonian structures, and further generalised the Lenard scheme
\cite{mr94j:58081}.

In this paper, we restrict to weakly nonlocal differential operators. Without using
Dirac structures, we prove that for compatible Hamiltonian and symplectic operators,
hierarchies of infinitely many commuting local symmetries and conservation laws can
be generated under some easily verified conditions. Their nonlocal terms suggest the
starting points of the Lenard scheme. We treat the case when the generating Nijenhuis
operators are weakly nonlocal in Theorem \ref{th1} in section \ref{sec31} and the
case when the generating Nijenhuis operators may not be weakly nonlocal in Theorem
\ref{th2} in section \ref{sec32}.

We apply these results to study the algebraic structures of the following system
\begin{eqnarray}\label{sys}
\left\{\begin{array}{ll} \phi_{1,t}=\phi_{1,xx}+2\phi_{2,xx} + 2 \phi_{1,x}
\phi_{2,x}
+\phi_{1,x}^2 +3 e^{2 \phi_{2}} -3 e^{-2 \phi_1-2\phi_2}\\
\phi_{2,t}=-2\phi_{1,xx}-\phi_{2,xx} - 2 \phi_{1,x} \phi_{2,x} -\phi_{2,x}^2 -3 e^{2
\phi_{1}} +3 e^{-2 \phi_1-2\phi_2}
\end{array}\right. .
\end{eqnarray}
We construct a recursion operator of system (\ref{sys}) from its Lax representation
by applying the idea in \cite{mr2001h:37146} for Lax pairs invariant under reduction
groups \cite{mik79,mik81,LM04,LM05} and prove it is a Nijenhuis operator in section
\ref{sec42}. In section \ref{sec5}, we show this system is a bi-Hamiltonian system by
constructing a Hamiltonian operator and a symplectic operator. Rather surprisingly,
the product of its weakly nonlocal Hamiltonian and symplectic operators gives rise to
the square of the Nijenhuis recursion operator. The Nijenhuis operator itself does
not possess such a decomposition. This phenomenon is very rare. The only known example
to me in the literature is the system of gas dynamics \cite{mr90b:58085}, which is of
hydrodynamical type and where the operators are local.

System (\ref{sys}) corresponds to the two dimensional periodic Volterra chain
\begin{eqnarray}\label{2+1}
\left\{ \begin{array}{l}\phi_{n,t}= \theta_{n,x} + \theta_n \phi_{n,x}
-e^{2 \phi_{n-1}}+ e^{2 \phi_{n+1}}, \\
\theta_{n+1}-\theta_n+ \phi_{n+1,x}+\phi_{n,x}=0 , \end{array}\right.\quad
\phi_{n+N}=\phi_n, \quad \sum_{n=1}^N \phi_n=0 .
\end{eqnarray}
with period $N=3$ \cite{mik79,mik81}.  It has also appeared in the classification of
integrable systems of nonlinear Schr{\"o}dinger type \cite{mr89g:58092}.

The two dimensional Volterra system (\ref{2+1}) can be viewed as a discretisation of
the Kadomtsev-Petviashvili equation. Indeed, in the limit $N\to \infty$,
\[\phi_n(x,t)=h^2 u(\xi,\eta,\tau),\quad h=N^{-1}, \]
\[ \tau=h^3 t, \quad \xi=n h+4ht,\quad \eta=h^2 x \]
 system (\ref{2+1}) goes to
\[ u_\tau=\frac{2}{3}u_{\xi\xi\xi}+8uu_\xi-2D_\xi^{-1}u_{\eta\eta}+O(h^2).\]
For integrable equations in the $2+1$-dimension, their recursion operators are no
longer pseudo-differential operators \cite{FS88c,FS88b,mr94h:35241}. To study the
family of discrete integrable system (\ref{2+1}) for fixed N, we wish to shed some
light on this issue. The exact solutions of system (\ref{2+1}) have much in common
with $2+1$-dimensional integrable equations, which have recently been investigated by
Bury and Mikhailov \cite{bm}.

\section{Definitions of geometric operators}\label{sec2}
In this section, we sketch the basic definitions of Hamiltonian, symplectic and
Nijenhuis operators following \cite{GD79, mr94j:58081, wang98}. We begin with
the construction of a complex of variational calculus.
\subsection{Complex of variational calculus}
Let $x,t$ be the independent variables and $u$ be a $N$-dimensional vector-valued
dependent variable, where $N\in \mathbb{N}$ is finite. All smooth functions depending
on $u$ and $x$-derivatives of $u$ up to some finite, but unspecified order form a
differential ring $\A$ with total $x$-derivation
$$D_x=\sum_{k=0}^\infty u_{k+1} \cdot \pderiv{}{u_k}, \quad \mbox{where}
\ u_k=\partial_x^k u. $$ Here $\cdot$ denotes the inner product of vectors. The
highest order of $x$-derivative we refer to the order of a given function. For any
element $g\in \A$, we define an equivalence class (or a functional) $\int\! g$ by
saying that $g$ and $h$ are equivalent if and only if \(g-h\in \im D_x\). The space
of functionals, denoted by $\A'$, does not inherit the ring structure from $\A$.

Any derivation $\partial $ on the ring $\A$ can be written as $\sum_{k=0}^\infty h_{k
}\cdot \pderiv{}{u_k}$, where $h_k$ is an $N$-dimensional vector field with entries
from $\A$. We denote the space of such vector fields as $h_k$ by $\A^N$. The
derivation $\partial$ is uniquely defined by its action on the dependent variable $u$
and its $x$-derivatives. The derivation commuting with $D_x$ can be recovered from
its action on the dependent variable, that is, $h_0$ since we have $h_k=D_x^k h_0$.
This is known as an evolutionary vector field. Let $\lieh$ denote the space of all
such $h_0$. For any $P\in \lieh$, there is a unique derivation
$\partial_P=\sum_{k=0}^\infty D_x^k P \cdot \pderiv{}{u_k} $. The natural commutator
of derivations leads to the Lie bracket on $\lieh$:
\begin{eqnarray}\label{lie}
\begin{array}{c}[P, \ Q]=D_Q[P]-D_P[Q],\quad P, Q \in \eta,\end{array}
\end{eqnarray}
where $D_Q=\sum_{i=0}^\infty\frac{\partial Q}{\partial u_i} \cdot D_x^i$ is the
Fr{\'e}chet derivative of $Q$.

We define the action of any element $P\in \lieh$ on $\int g \in \A'$ as follows:
\begin{eqnarray}\label{act}
\begin{array}{c}P \int g =\int \partial_P(g)=\int \sum_{k=0}^\infty D_x^k P
\cdot \frac{\partial
g}{\partial u_k}=\int D_g[P].\end{array}
\end{eqnarray}
Direct computation shows that such an action is a representation of the Lie algebra
$\lieh$. Having a representation space of Lie algebra $\lieh$, we can build an
associated Lie algebra complex. This complex is called the complex of variational
calculus. Let us give the first few steps since we do not need the general theory.

We denote the space of functional $n$-forms by $\Omega^n$ starting with
$\Omega^0=\A'$. Now we consider the space $\Omega^1$.  For any vertical 1-form on the
ring $\A$, i.e., $\omega=\sum_{k=0}^\infty h^k \cdot \d u_k$, where $h^k\in \A^N$,
there is a natural non-degenerate pairing with the derivations $\partial_P$:
\begin{eqnarray}\label{pairing}
\begin{array}{c}<\omega, \ P>=\int \sum_{k=0}^\infty h^k \cdot D_x^k P =
\int \left(\sum_{k=0}^\infty (-D_x)^k h^k \right) \cdot  P \ .\end{array}
\end{eqnarray}
This can be viewed as the pairing of $1$-forms of the form $\xi\ \d u$ with $P\in
\lieh$. Thus any element of $\Omega^1$ is completely defined by $\xi\in \A^N$. For a
given $\omega$, we have $\xi=\sum_{k=0}^\infty (-D_x)^k h^k$.

The pairing between Lie algebra $\lieh$ and $1$-forms $\Omega^1$ allows us to give
the definition of (formal) adjoint operators to linear (pseudo)-differential
operators \cite{mr94g:58260}.
\begin{Def} Given a linear operator ${\cal S}: \lieh \rightarrow  \Omega^1$,
we call the operator
${\cal S}^{\star}: \lieh \rightarrow \Omega^1$ the adjoint operator of ${\cal S}$ if
$<{\cal S} P_1, \ P_2>=<{\cal S}^{\star} P_2, \ P_1>$, where $P_i\in \lieh$ for
$i=1,\ 2$.
\end{Def}
Similarly, we can define the adjoint operator for an operator mapping from $\Omega^1$
to $\lieh$, from $\lieh$ to $\lieh$ or from $\Omega^1$ to $\Omega^1$.

The variational derivative associates with each functional $\int\!\! g\in \A'$ its
Euler-Lagrange expression $\delta(\int\!\! g) \in \Omega^1$ defined so that
\begin{eqnarray}\label{euler}
\begin{array}{c}<\delta(\int \!g),\ P>=(\d \int\! g) (P)=P \int \! g=
<\sum_{k=0}^\infty (-D_x)^k \frac{\partial g}{\partial u_k},\ P> \ ,\end{array}
\end{eqnarray}
where $\d: \Omega^n\rightarrow \Omega^{n+1}$ is a coboundary operator. Due to the
non-degeneracy of the pairing (\ref{pairing}), we have $\delta(\int\!
g)=\sum_{k=0}^\infty (-D_x)^k \frac{\partial g}{\partial u_k}\in \Omega^1.$ In the
literature one often uses $E$ referring to the Euler
operator instead of $\delta$. For any $\xi\in \Omega^1$, by direct calculation we obtain $\d \xi =
D_{\xi} - D_{\xi}^{\star}$. We say that the $1$-form $\xi$ is closed if $\d \xi=0$.

Finally, we give the formulas of Lie derivatives  along any $K\in \lieh$ using
Fr{\'e}chet derivatives, cf.  \cite{mr94j:58081} for the details.
\begin{Def}\label{def0}
Let $L_K$ denote the Lie derivative along $K\in \lieh$. We have
\begin{equation*}
\begin{array}{l} L_K\!\int\!\! g=\int\! D_{g}[K] \ \ \mbox{for} \ \ \int\!\! g\in\A';\\
L_K\! h=[K, h] \ \ \mbox{for} \ \ h\in\lieh;\\
L_K\!\xi =D_{\xi}[K] +D_K^{\star}(\xi) \ \ \mbox{for} \ \ \xi\in\Omega^1;\\
L_K\! \Re=D_{\Re}[K] -D_K \Re+\Re D_K \ \ \mbox{for} \ \ \Re: \lieh \rightarrow
\lieh;\\
L_K\! \cH=D_{\cH}[K] -D_K \cH-\cH D_K^{\star}\ \
\mbox{for} \ \  \cH: \Omega^1 \rightarrow \lieh;\\
L_K \!\J=D_{\J}[K] +D_K^{\star} \J+\J D_K\ \ \mbox{for} \ \  \J: \lieh \rightarrow
\Omega^1.\end{array}
\end{equation*}
\end{Def}
In this complex we can identify all the important concepts in the study of integrable
systems such as symmetries, cosymmetries, conservation laws and recursion operators.
They are all characterised by the vanishing of the Lie derivatives with respect to a
given evolution equation. This will be discussed further in section \ref{sec3}.

\subsection{Symplectic, Hamiltonian and Nijenhuis operators}
\begin{Def} A linear operator ${\cal S}: \lieh \rightarrow \Omega^1$ (or
$ \Omega^1 \rightarrow  \lieh $) is anti-symmetric if $ {\cal S}= -{\cal S}^{\star}$.
\end{Def}

Given an anti-symmetric operator $\J: \lieh\rightarrow \Omega^1$, there is an
anti-symmetric $2$-form associated with it. Namely,
\begin{eqnarray}\label{2fom}
\begin{array}{c}\omega(P,Q)=<\J(P),Q>=-<\J(Q),P>=-\omega(Q,P), \quad P,Q\in \lieh .
\end{array}
\end{eqnarray}
Here the functional $2$-form $\omega$ has the canonical form \cite{mr94g:58260}
\begin{eqnarray}\label{2fomw}
\omega=\frac{1}{2} \int\!\! \d u \wedge \J \d u.
\end{eqnarray}
\begin{Def}
An operator $\J: \lieh\rightarrow \Omega^1$ is called symplectic if and only if the
anti-symmetric $2$-form (\ref{2fomw}) is closed, i.e., $\d\omega=0$.
\end{Def}
The symplecticity condition $\d\omega=0$ can be presented in several explicit and
equivalent versions. The details can be found in Theorem 6.1 in \cite{mr94j:58081}.


For an anti-symmetric operator $\cH: \Omega^1\rightarrow \lieh$, we can define a
Poisson bracket of two functionals
\begin{eqnarray}\label{poi}
\begin{array}{l}\left\{ \int\!\! f,\ \int\!\! g\right\}=<\delta( f), \cH \delta( g)> .
\end{array}
\end{eqnarray}
\begin{Def}
The operator $\cH$ is Hamiltonian if the Poisson bracket defined by (\ref{poi}) is
anti-symmetric and satisfies the Jacobi identity
$$\begin{array}{l}\left\{\left\{ \int\!\! f, \int\!\! g \right\},\ \int\!\! h \right\}+
\left\{\left\{ \int\!\! g, \int\!\! h \right\},\ \int\!\! f \right\}+\left\{\left\{
\int\!\! h, \int\!\! f \right\},\ \int\!\! g \right\}=0 .\end{array}$$
\end{Def}
For the Jacobi identity, there are several equivalent formulas given in
\cite{mr94j:58081} (see Theorem 5.1). In \cite{mr94g:58260} (see p. $443$), it was
formulated as the vanishing of the functional tri-vector:  $ \int \theta \wedge
D_{\cH} [\cH \theta] \wedge \theta =0 $, which makes it feasible to check.

The Jacobi identity is a quadratic relation of the operator $\cH$. In general, the
linear combination of two Hamiltonian operators is no longer Hamiltonian. If it is,
we say that two such Hamiltonian operators form a Hamiltonian pair. Hamiltonian pairs
play an important role in the theory of integrability. They naturally generate
Nijenhuis operators.
\begin{Def} A linear operator $\Re: \lieh\rightarrow \lieh$ is called a Nijenhuis
operator if it satisfies
\begin{eqnarray}\label{Nijen}
\begin{array}{l}[\Re P, \Re Q]-\Re[\Re P, Q]-\Re[P, \Re Q]+\Re^2[P, Q]=0,
\quad P, Q\in \lieh.\end{array}
\end{eqnarray}
\end{Def}
Using the definition of Lie bracket (\ref{lie}), formula (\ref{Nijen}) is equivalent
to
\begin{eqnarray}\label{Nijen1}
\begin{array}{l}L_{\Re P} \Re =\Re L_P \Re .\end{array}
\end{eqnarray}
An equivalent formulation is: $D_{\Re}[\Re P](Q)-\Re D_{\Re}[P] (Q)$ is symmetric
with respect to $P$ and $Q$, cf. \cite{mr84j:58046}.

The properties of Nijenhuis operators \cite{mr94j:58081} provide us with the explanation
how the infinitely many commuting symmetries and conservation laws
of integrable equations arise.
\section{Lenard Scheme of Integrability}\label{sec23} The Lenard scheme
was first used to generate the KdV hierarchy \cite{ps05}.
After the discovery of the interrelations between Hamiltonian
pairs and Nijenhuis operators \cite{GD79, mr84j:58046} it was applied to bi-Hamiltonian systems.
In 1987, Dorfman (\cite{dorf87}) introduced the concept of Dirac structures into the field of soliton
theory in order to deal with nonlocal terms in the operators.

In this section, we consider {\em weakly
nonlocal} \cite{MaN01} Hamiltonian and symplectic operators, i.e.,
pseudo-differential operators with only a finite
number of nonlocal terms of the form $ P \otimes D_x^{-1} Q$, where $P$ and $Q $ are
in the Lie algebra $\lieh$ for Hamiltonian operators and in the space of $1$-forms
$\Omega^1$ for symplectic operators. We prove that for compatible weakly nonlocal
Hamiltonian and symplectic operators, hierarchies of commuting local symmetries and
cosymmetries can be generated under some easily verified conditions without using
Dirac structures.
This is independent of the generating Nijenhuis operators being weakly
nonlocal or not. Their nonlocal terms suggest the starting points of the Lenard
scheme.

\begin{Def} A Hamiltonian operator $\cH:\Omega^1 \rightarrow \lieh$ and a symplectic
operator $\J: \lieh\rightarrow \Omega^1$ are compatible if $\Re=\cH\J$ is a Nijenhuis
operator.
\end{Def}
We assume without loss of generality the nonlocal terms of a weakly nonlocal
Hamiltonian operator $\cH:\Omega^1 \rightarrow \lieh$ are of the form \cite{Mal2}
\begin{equation}\label{H}
\begin{array}{l}\sum_{j=1}^m \epsilon_j P_j \otimes D_x^{-1} P_j, \quad \mbox{where} \ \
\epsilon_j\in \{-1,0,1\}\ \ \mbox{and} \ \  P_j\in \lieh
\end{array}\end{equation}
and those of a symplectic operator $\J: \lieh\rightarrow \Omega^1$ are of the form
\begin{equation}\label{J}
\begin{array}{l}\sum_{k=1}^n {\tilde \epsilon}_k  \gamma_k \otimes D_x^{-1}
\gamma_k, \quad \mbox{where}
 \ \ {\tilde \epsilon}_k\in \{-1,0,1\},\  \gamma_k\in \Omega^1 \ \ \mbox{and} \ \
D_{\gamma_k}=D_{\gamma_k}^{\star}
\end{array}\end{equation}
with the convention that if $\epsilon_j=0$ or ${\tilde \epsilon}_k=0$, we take $P_j=0$ or
$\gamma_k=0$. Here $\otimes$ denotes the matrix product of two vectors ($N\times 1$
column matrices), producing a $N \times N$ matrix.

For a given weakly nonlocal operator $\cal S$, its highest power of $D_x$
is the order of an operator. We say that the operator ${\cal S}$ is
degenerate if there exists a non-zero weakly nonlocal operator ${\cal T} $ such that
${\cal S T}=0$. Otherwise, we say that the operator ${\cal S}$ is non-degenerate.

Notice that the pairing between $P_j\in \lieh$ and $\gamma_k$ appears in the computation
of $\cH \J$. It is important to determine whether the pairing is zero or not. The
pairings being zero implies that operator $\cH \J$ is again weakly nonlocal.
\begin{Lem}\label{lem3}
For any $Q\in \lieh$ and $\xi\in \Omega^1$ if $D_{\xi}=D_{\xi}^{\star}$ then
$L_{Q}\xi=\delta\!<\!\!\xi,\ Q\!\!>$.
\end{Lem}
{\bf Proof}. Indeed, for any $P\in \lieh$ we have
\begin{eqnarray*}
\begin{array}{l}\quad <\delta\! <\!\!\xi,\ Q\!\!>, P> =<D_{\xi}[P], Q>
+<\xi, D_Q[P]>=<D_{\xi}^{\star} Q+D_Q^{\star}\xi,\ P>\\
\ =<D_{\xi} Q+D_Q^{\star}\xi,\ P>=<L_Q\xi,P>.
\end{array}
\end{eqnarray*}
Since the pairing is non-degenerate, we obtain $L_{Q}\xi=\delta\!<\!\!\xi,\ Q\!\!>$.
\hfill $\blacksquare$
\begin{Pro} Let the nonlocal terms of operators, $\cH$ and $\J$ be of the form as
(\ref{H}) and (\ref{J}). Assume that $P_j$, $j=1,\cdots, m$ and $\gamma_k$,
$k=1,\cdots, n$ are linear independent over $\mathbb{C}$, respectively. If $L_{P_j}
\J=L_{P_j}\cH=0$, then there exists anti-symmetric constant $m\times m$ matrix
$A^{(j)}$ and $n \times n$ matrix $B^{(j)}$ such that
$$ \epsilon_i L_{P_j} P_i=\sum_{k=1}^m P_k A^{(j)}_{ki} \quad \mbox{and} \quad
\tilde \epsilon_i L_{P_j} \gamma_i=\sum_{k=1}^n \gamma_k B^{(j)}_{ki}. $$
\end{Pro}
{\bf Proof}. The assumption $L_{P_j} \J=L_{P_j}\cH=0$ implies that
\begin{equation*}\begin{array}{l}\sum_{i=1}^m \left(\epsilon_i
L_{P_j} P_i \otimes D_x^{-1} P_i +P_i \otimes D_x^{-1} \epsilon_i L_{P_j} P_i\right)
=0;\\
\sum_{i=1}^n \left({\tilde \epsilon}_i L_{P_j} \gamma_i \otimes D_x^{-1} \gamma_i +
\gamma_i \otimes D_x^{-1}  {\tilde \epsilon}_i L_{P_j} \gamma_i  \right)=0.
\end{array}\end{equation*}
Applying Theorem \ref{apth}, we obtain the results as stated. \hfill $\blacksquare$

If for all $j=1,\cdots, m$ the matrices $B^{(j)}$ are zero, the operator $\cH \J$ is weakly
nonlocal since $\delta< \gamma_i, P_j>=L_{P_j} \gamma_i=0$ according to Lemma
\ref{lem3}. A lot of work has been done for this case, e.g. \cite{mr1974732,serg5}.
We give the Lenard scheme including all the starting points in section \ref{sec31}.
If there exists $B^{(j)}\neq 0$, the operator $\cH \J$ is no longer weakly nonlocal.
The locality in this case has not been answered so far. In section \ref{sec32}, we
tackle this problem and work out concrete examples.

\subsection{Case I:  operator $\cH\J$ is weakly nonlocal}\label{sec31}
\begin{The}\label{th1}
Let $\cH$ and $\J$ be compatible Hamiltonian and symplectic operators, as defined above.
Assume that $L_{P_j}P_l=L_{P_j} \gamma_k=L_{P_j} \J=L_{P_j}\cH =L_{\cH
\gamma_k} \gamma_s=0$ and $\J \cH \gamma_k$ is closed, where $j, l=1, \cdots, m$ and
$k, s=1,\cdots, n$. Then for all $i, i_1=0, 1, 2 , \cdots ,$
\begin{enumerate}
\item $\xi_k^i=(\J\cH)^i \gamma_k \in \Omega^1 $ are closed $1$-forms and
$h_k^{i}=\cH \xi_k^i \in \lieh$ commute;
\item $p_j^i=(\cH \J)^i P_j \in \lieh$ commute and $\zeta_j^i=\J p_j^i\in \Omega^1$ are
closed $1$-forms;
\item vector fields $h_k^{i}$ and $p_j^{i_1}$ commute for all $j=1, \cdots, m$,
$k=1,\cdots, n$.
\end{enumerate}
If there exist $f_k^i$ and $g_j^i$ such that $\xi_k^i=\delta f_k^i$ and
$\zeta_j^i=\delta g_j^i$, then all $h_k^i$ and $p_j^i$ are Hamiltonian vector fields
and their Hamiltonian are in involution.
\end{The}
Before we proceed with the proof, we first give a few lemmas. Part of Lemma \ref{lem1} and
Lemma \ref{lem2} have been formulated and proved in abstract manner in
\cite{mr94j:58081} (see Proposition 2.4 and 2.8). Here we give a straightforward
proof.

\begin{Lem}\label{lem1}
Let $\J$ be a symplectic operator. For $P\in \lieh$ such that $\J P \in \Omega^1$,
then $\J P$ is closed if and only if $L_{P}\J=0$.
\end{Lem}
{\bf Proof}. We know $D_{\J P}=D_{\J P}^{\star}$ if and only if for any $Q, H\in
\lieh$,
\begin{equation*}
\begin{array}{l} <D_{\J P}[Q], H>=<D_{\J P}^{\star}(H), Q>=<D_{\J P}[H],Q>,
\end{array}
\end{equation*}
that is,
\begin{equation*}
\begin{array}{l} 0=<D_{\J P}[Q], H>-<D_{\J P}[H],Q>\\
\quad =<D_{\J}[Q](P),H>+<\J D_P[Q],H>-<D_{\J}[H](P),Q>-<\J D_P[H],Q>\\
\quad =-<D_{\J}[P](H),Q>-<D_P^{\star} \J [H],Q>-<\J D_P[H],Q>,
\end{array}
\end{equation*}
where we used the fact that $\J$ is a symplectic operator. This leads to $$
D_{\J}[P]+D_P^{\star} \J+\J D_P=0.$$ From Definition \ref{def0}, it follows that
$L_{P}\J=0$.  \hfill $\blacksquare$
\begin{Lem}\label{lem2}
Let $\cH$ be a Hamiltonian operator. If $\cH  \xi=P$ for some $\xi\in \Omega^1$ and
$D_{\xi}=D_{\xi}^{\star}$, then $L_{P}\cH=0$.
\end{Lem}
{\bf Proof}. We know that $L_{P} \cH=D_{\cH}[P]-D_{P}\cH -\cH D_P^{\star}$. For any
$p, q \in \Omega^1$, we compute
\begin{equation*}
\begin{array}{l} \quad <(L_{P}\cH) (p), q>=<D_{\cH}[\cH \xi](p),q>
-<D_{\cH \xi}[\cH p], q> -<\cH D_{\cH \xi}^{\star}(p),q>\\
=-<D_{\cH}[\cH q](\xi),p>-<\cH D_{\xi}[\cH p], q>+< D_{\cH \xi}[\cH q],p>=0,
\end{array}
\end{equation*}
where we used $D_{\xi}=D_{\xi}^{\star}$ and $\cH$ being a Hamiltonian operator. This
leads to $L_{P}\cH=0$.  \hfill $\blacksquare$

{\bf Proof of Theorem \ref{th1}}. The assumption that $L_{P_j} \gamma_k=0$ and the
closedness of  $\xi$ leads to $\delta\!<\!\!\gamma_k,\ P_j\!\!>=0$. Thus the operator
$\Re=\cH \J$ is weakly nonlocal with nonlocal terms as follows:
\begin{eqnarray*}
\begin{array}{l}\sum_{j=1}^m \epsilon_j P_j \otimes D_x^{-1} (\J P_j)
+\sum_{k=1}^n {\tilde \epsilon}_k (\cH \gamma_k) \otimes D_x^{-1} \gamma_k
\end{array}.
\end{eqnarray*}
Besides, $\Re$ is Nijenhuis since $\cH$ and $\J$ are compatible.

We first check the conditions of statement \ref{form} in Lemma \ref{lem4} in
the Appendix. For the first statement in the theorem, we only need to check $L_{P_j}
\Re=L_{\cH \gamma_k}\Re=0$. We know  $$L_{P_j} \Re=L_{P_j}(\cH \J)=L_{P_j}(\cH) \J+
\cH L_{P_j}(\J)=0.$$ Now we show $L_{\cH \gamma_k}\Re=0$. Since $L_{P_j} \gamma_k=0$,
so $\cH \gamma_k$ is local. Together with $D_{\gamma_k}=D_{\gamma_k}^{\star}$, we
have $L_{\cH \gamma_k} \cH=0$ by Lemma \ref{lem2}. The assumption that $\J \cH
\gamma_k$, which is local due to $L_{\cH \gamma_k} \gamma_i=0$, is closed leads to
$L_{\cH \gamma_k} \J=0$ from Lemma \ref{lem1}. So $L_{\cH \gamma_k}\Re=L_{\cH
\gamma_k}(\cH \J)=0$. Thus we prove that $1$-forms $\xi_k^i=(\J\cH)^i \gamma_k $ are
local and closed.

For the second statement in the theorem, we need to show $L_{P_j}(\J P_l)=L_{\cH
\gamma_k} (\J P_j)=0$ and $\J\cH \J P_j$ is closed.

It is clear that $L_{P_j}(\J P_l) =L_{P_j}(\J) (P_l)+\J L_{P_j}( P_l)=0$ by the
assumptions. We also know
\begin{equation}\label{com}
L_{P_j} (\cH \gamma_k)=L_{P_j} \cH (\gamma_k)+ \cH (L_{P_j}  \gamma_k)=0.
\end{equation}
So $L_{\cH \gamma_k} (\J P_j)=L_{\cH \gamma_k} (\J) (P_j)+ \J L_{\cH \gamma_k} (
P_j)=0$. From Lemma \ref{lem1}, $\J\cH \J P_j$ being closed is equivalent to $L_{\cH
\J P_j} \J=L_{\Re P_j} \J=0$. It follows from Lemma \ref{lem1} and $L_{P_j}\J=0$ that
$\J P_j$ is closed. We also know $\Re P_j$ is local since we have proved $L_{P_j}(\J
P_l)=0$. Thus $L_{\Re P_j} \cH=0$ according to Lemma \ref{lem2}. Since $\Re$ is
Nijenhuis, we have $L_{\Re P_j} \Re=\Re L_{P_j} \Re=0$, that is, $0=(L_{\Re
P_j}\cH)\J =\cH L_{\Re P_j}\J=\cH L_{\Re P_j}\J$ implying $L_{\Re P_j}\J=0$. We can
now draw the conclusion that $1$-forms $\zeta_j^i=\J p_j^i$ are local and closed.

Next the conditions of statement \ref{flow} in Lemma \ref{lem4} in Appendix are
satisfied when we take into account what we have proved. Thus we can conclude for
fixed $k$ and $j$, vector fields $h_k^{i}=\cH \xi_k^i$ commute and $p_j^i=(\cH \J)^i
P_j$ commute.

The third statement in the theorem follows from the fact $\Re$ is Nijenhuis and
$$L_{P_j} P_l=L_{\cH \gamma_k} (\cH \gamma_s)=L_{P_j} (\cH \gamma_k)=0$$ for all
$j,\ l=1, \cdots, m$ and $k,\ s=1,\cdots, n$.

Finally, we prove  $\left\{f_k^i, \ g_j^{i_1}\right\}=0$ if $f_k^i$ and $g_j^{i_1}$
exist. The other cases $\left\{f_k^i, \ f_{k_1}^{i_1}\right\}=0$ and
$\left\{g_{j_1}^i, \ g_j^{i_1}\right\}=0$ can be proved in the same way. We have
$$\left\{f_k^i, \ g_j^{i_1}\right\}=<\xi_k^i, \cH \zeta_j^{i_1}>=<(\J \cH)^i \gamma_k,
 (\cH \J)^{i_1+1}P_j>=<\gamma_k,\ \Re^{i_1+i+1} P_j>$$
and
\begin{equation*}
\begin{array}{l}\quad \delta <\gamma_k,\ \Re^{s+1} P_j>=L_{\Re^{s+1} P_j}\gamma_k=
\delta <\J \cH \gamma_k,\ \Re^{s} P_j> =L_{\Re^{s} P_j} (\J \cH \gamma_k)\\
=(L_{\Re^{s} P_j} (\J \cH)) \gamma_k+\J \cH L_{\Re^{s} P_j} (\gamma_k)=\J \cH
L_{\Re^{s} P_j} (\gamma_k). \end{array}
\end{equation*}
By induction, we obtain $L_{\Re^{s+1} P_j}\gamma_k=\delta <\gamma_k,\ \Re^{s+1}
P_j>=0$ implying $\left\{f_k^i, \ g_j^{i_1}\right\}=0$.  By now, we complete the
proof. \hfill $\blacksquare$

\begin{figure}[ht]
\begin{eqnarray*}
\begin{array}{ccc}
\begin{pspicture}(0,0)(5,5)
\rput(2.9,4.5){$\xi_k^0=\gamma_k$} \rput(0,3.8){$h_k^0$}\rput(1.4,4.5){$\cH$}
\psline{->}(2.3,4.5)(0.2,3.8) \rput(2.5,3.1){$\xi_k^1$}
\rput(0,2.4){$h_k^1$}\rput(1.4,3.05){$\cH$} \psline{->}(2.3,3.1)(0.2,2.4)
\rput(2.5,1.7){$\xi_k^2$} \rput(0,1.0){$h_k^2$}\rput(1.4,1.58){$\cH$}
\psline{->}(2.36,1.6)(0.2,1.0) \psline{->}(0,3.8)(2.28,3.2)\rput(1.4,3.66){$\J$}
\psline{->}(0,2.4)(2.28,1.8)\rput(1.4,2.26){$\J$}\psline{->}(0,1.0)(2.3,0.4)
\rput(1.4,0.86){$\J$}\rput(0,0){$\cdots$}\rput(1.2,0){$\cdots$}\rput(2.4,0){$\cdots$}
\end{pspicture}& \qquad &
\begin{pspicture}(0,0)(5,5)
\rput(0,4.5){$P_j=p_j^0$}
\psline{->}(0.5,4.5)(2.8,3.9)\rput(1.8,4.4){$\J$}\rput(3,3.8){$\zeta_j^0$}
\psline{->}(2.9,3.8)(0.65,3.1)\rput(1.8,3.71){$\cH$}  \rput(0.5,3.1){$p_j^1$}
\psline{->}(0.5,3.1)(2.8,2.5)\rput(1.8,3.0){$\J$}\rput(3,2.4){$\zeta_j^1$}
\psline{->}(2.9,2.4)(0.65,1.7)\rput(1.8,2.31){$\cH$}  \rput(0.5,1.7){$p_j^2$}
\psline{->}(0.5,1.7)(2.8,1.1)\rput(1.8,1.6){$\J$}\rput(3,1.0){$\zeta_j^2$}
\psline{->}(2.9,1.0)(0.65,0.3)\rput(1.8,0.91){$\cH$}
\rput(0.5,0){$\cdots$}\rput(1.7,0){$\cdots$}\rput(2.9,0){$\cdots$}
\end{pspicture}
\end{array}
\end{eqnarray*}
\caption{Lenard scheme for compatible Hamiltonian and symplectic\newline operators
when the generating Nijenhuis operator is weakly nonlocal}\label{fig0}
\end{figure}
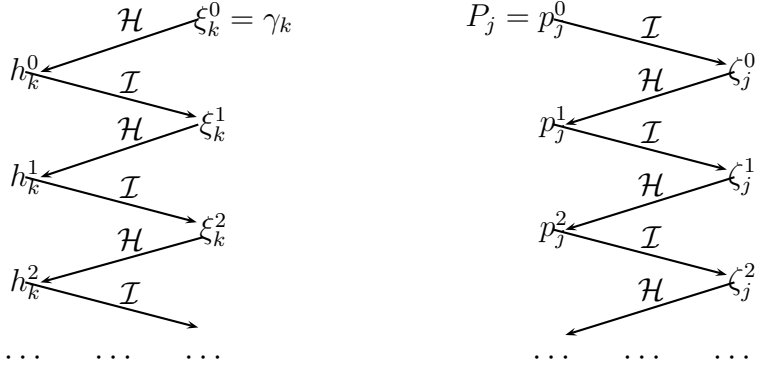

This theorem gives rise to the Lenard scheme as shown in Figure \ref{fig0} for fixed
$k$ and $j$.

In fact, such a scheme has been implicitly used for some integrable equations including
the new systems in \cite{mnw07}. Since we have, for $P, Q \in \lieh$ or $P, Q \in
\Omega^1$,
$$P\otimes D_x^{-1} Q +Q \otimes D_x^{-1} P=\frac{1}{2} (P+Q) \otimes D_x^{-1}
(P+Q)-\frac{1}{2} (P-Q) \otimes D_x^{-1} (P-Q),$$ we can easily adapt Theorem
\ref{th1} in terms of $P$ and $Q$ instead of $P+Q$ and $P-Q$ since all the operations
involved in the theorem such as Lie derivatives and Poisson bracket are linear.
\begin{Ex}
The Sawada-Kotera equation
\begin{eqnarray}\label{sk}
u_t=u_{5x}+5 u u_{3x}+5 u_x u_{2x}+5 u^2 u_x
\end{eqnarray}
has a Hamiltonian operator $$\cH=D_x^3+2u D_x+2 D_x u$$  and a compatible symplectic
operator \cite{mr83d:58031} $$\J=D_x^3+u D_x+D_x u +(u_{xx}+\frac{u^2}{2})
D_x^{-1}+D_x^{-1} (u_{xx}+\frac{u^2}{2}). $$ Starting from both $1$ and
$u_{xx}+\frac{u^2}{2}$ as $\xi_1^0$ and $\xi_2^0$ in Figure \ref{fig0}, we generate
the hierarchies of symmetries and cosymmetries of the Sawada-Kotera equation.
\end{Ex}

\subsection{Case II:  operator $\cH\J$ is not weakly nonlocal}\label{sec32}
We start with a known example: although both the Hamiltonian and the symplectic operator are
weakly nonlocal, the operator $\cH \J$ is {\em not} weakly nonlocal.
\begin{Ex}\label{eg1} The two-component system
\begin{equation}\label{SS}\left\{\begin{array}{l}u_t=u_{xxx}+9 u u_x v+3 u^2 v_x\\
v_t=v_{xxx}+9 u v v_x+3v^2 u_x\end{array}\right.\end{equation} possesses a
Hamiltonian operator
\begin{equation*}
\cH=\left(\begin{array}{cc}-u D_x^{-1} u & D_x+u D_x^{-1} v\\
D_x+v D_x^{-1} u & -v D_x^{-1} v\end{array}\right)
=\left(\begin{array}{cc}0 & D_x\\
D_x & 0\end{array}\right)-\left(\!\!\begin{array}{c}u \\
-v\end{array}\!\!\right) D_x^{-1}\left(\begin{array}{ll}u & -v\end{array}\right)
\end{equation*}
and a symplectic operator
\begin{eqnarray*}
&&\quad\J=\left(\begin{array}{cc}3 v D_x^{-1} v & D_x+2 u D_x^{-1} v+3v D_x^{-1} u\\
D_x+2 v D_x^{-1} u +3 u D_x^{-1} v& 3 u D_x^{-1} u\end{array}\right)\\
&&=\left(\!\!\begin{array}{cc}0 & D_x\\D_x & 0\end{array}\!\!\right)+3\left(\!\!
\begin{array}{c}v \\
u\end{array}\!\!\right) D_x^{-1}\left(\!\!\begin{array}{ll}v &
u\end{array}\!\!\right)+\left(\!\!\begin{array}{c}u \\
v\end{array}\!\!\right) D_x^{-1}\left(\!\!\begin{array}{ll}u&
v\end{array}\!\!\right)-\left(\!\!\begin{array}{c}u \\
-v\end{array}\!\!\right) D_x^{-1}\left(\!\!\begin{array}{ll}u &
-v\end{array}\!\!\right).
\end{eqnarray*}
These compatible Hamiltonian and symplectic structures of system (\ref{SS})
first appeared in \cite{serg7b}, although the given symplectic operator was
incorrect as stated there.
\end{Ex}
Using the notations in (\ref{H}) and (\ref{J}), we have
\begin{eqnarray}\label{eg1p}
P_1=\left(\!\!\begin{array}{c}u \\ -v\end{array}\!\!\right); \quad
\gamma_1=\left(\!\!\begin{array}{c}u \\ v\end{array}\!\!\right); \quad
\gamma_2=\left(\!\!\begin{array}{c}u \\ -v\end{array}\!\!\right); \quad
\gamma_3=\left(\!\!\begin{array}{c} \sqrt 3 v \\ \sqrt 3 u\end{array}\!\!\right)  .
\end{eqnarray}
Indeed,
\begin{eqnarray*}
\left(\!\!\begin{array}{ccc}L_{P_1} \gamma_1, &  -L_{P_1}\gamma_2, & L_{P_1} \gamma_3
\end{array}\!\!\right)=\left(\!\!\begin{array}{ccc} \gamma_1, &  \gamma_2, &  \gamma_3
\end{array}\!\!\right)\left(\!\!\begin{array}{ccc}0& -2 & 0\\
2& 0 &0\\ 0& 0 & 0 \end{array}\!\!\right) \neq 0
\end{eqnarray*}
and thus we can not present $\cH \J$ as a weakly nonlocal operator.
\begin{Def} We say $Q_1\equiv Q_2$, where $Q_1, Q_2 \in \lieh$, with respect to
$\{\beta_1, \cdots, \beta_n\}$, where $\beta_i\in \Omega^1$ if $L_{Q_1-Q_2}
\beta_i=0$ for $i=1, \cdots, n$.
\end{Def}
\begin{Lem}\label{lem5}
Let $\gamma_i$, $i=1,2,3$ and $P_1$ be defined as (\ref{eg1p}) in Example \ref{eg1}.
For any $Q\in \lieh$ and anti-symmetric $3\times 3$ constant matrix $A$, if $\tilde
\epsilon_i L_Q \gamma_i =\sum_{k=1}^3 \gamma_k A_{ki}$, then $Q\equiv
\mbox{Span}_{\bbbc}<P_1>$ with respect to $\{\gamma_1,\gamma_2, \gamma_3\}$.
\end{Lem}
{\bf Proof}. From $L_Q \gamma_1=A_{21}\gamma_2+A_{31} \gamma_3$, we obtain $Q \equiv
\frac{1}{2} \left(\!\!\begin{array}{c}A_{21} u+A_{31} \sqrt 3 v,\ A_{31} \sqrt 3 u
-A_{21} v \end{array}\!\!\right)^{\tr}$ with respect to $\gamma_1$. For such $Q$, it
follows $A_{32}=0$ from  $-L_{Q}\gamma_2=-A_{21} \gamma_1+A_{32} \gamma_3$ and
further $A_{31}=0$ from  $L_{Q}\gamma_3=-A_{31} \gamma_1$. Thus we have $Q\equiv
\frac{A_{21}}{2} \left(\!\!\begin{array}{c}u,\  -v \end{array}\!\!\right)^{\tr}$.
\hfill $\blacksquare$

This lemma is inspired by Theorem \ref{apth} in Appendix. The idea is to identify
$Q\in \lieh$ such that $ \tilde \epsilon_k L_{Q}\gamma_k=\sum_{i=1}^n \gamma_i
A_{ik}$ when the anti-symmetric $n\times n$ constant matrix $A\neq 0$ for given a set
of linear independent $\gamma_i$, $i=1,\cdots, n$ over $\bbbc$. Such an evolutionary
vector field is rather strict. Therefore, we introduce the following definition.
\begin{Def}\label{defp}
We say $\xi\in \Omega^1$ is {\em proper} for the operators $\cH$ and $\J$ if for
all $1\leq l \in \bbbn$, vectors $(\cH \J)^l \cH \xi$ have no intersection with
$\mbox{Span}_{\bbbc}<P_1, \cdots, P_n>$.
\end{Def}
\begin{The}\label{th2}
Let Hamiltonian and symplectic operators, $\cH$ and $\J$ with nonlocal terms being
(\ref{H}) and (\ref{J}), be compatible. Assume that
\begin{enumerate}
\item $\cH$ is non-degenerate;\label{con1}
\item $\gamma_k$ are linear independent over $\mathbb{C}$ for $k=1, \cdots,
n$;\label{con3}
\item $L_{P_j} \J=L_{P_j}\cH =0$ for $j=1,\cdots , m$;\label{con2}
\item For an anti-symmetric $n\times n$ constant matrix $A$ and $Q\in \lieh$, if
$ \tilde \epsilon_k L_{Q}\gamma_k=\sum_{i=1}^n \gamma_i A_{ik}$,  then $Q\equiv
\mbox{Span}_{\bbbc}<P_1, \cdots, P_n>$ with respect to $\{\gamma_1, \cdots,
\gamma_n\}$.  \label{con4}
\end{enumerate}
Then for a proper closed $1$-form $\xi^0$ satisfying $L_{P_j} \xi^0 =L_{\cH \xi^0}
\gamma_k=L_{\cH \xi^0} \xi^0=0$ such that $\J \cH \xi^0$ is closed, all
$\xi^i=(\J\cH)^i \xi^0 \in \Omega^1 $ are closed $1$-forms and $h^{i}=\cH \xi^i \in
\lieh$ commute for $i=0, 1, 2 , \cdots $. If, moreover, $\xi^i=\delta f^i$, then all
$h^i$ are Hamiltonian vector fields and their Hamiltonians are in involution.
\end{The}

{\bf Proof}. Since $\cH$ and $\J$ are compatible, we have that $\Re=\cH \J$ is Nijenhuis.
From the definition of $h^0$ and Lemma \ref{lem1} and \ref{lem2} this leads to
$L_{h^0}\cH=L_{h^0}\J=0$ and thus $L_{h^0}\Re=0$.  Therefore, under the assumptions,
if $h^i\in \lieh$ and $\xi^i\in \Omega^1$, i.e., local, then the $h^i$ commute, $L_{h^i}
\Re=0$ and the $\xi^i$ are closed. Thus we only need to show that $h^i$ and $\xi^i$ are local.

Assume that $h^{l-1}$ and $\xi^l$ are local and $L_{P_j} \xi^l=0$ for $l\geq 1$. We
show that $h^l$ and $\xi^{l+1}$ are local and $L_{P_j}\xi^{l+1}=0$ by induction. It
follows from Lemma \ref{lem3} that $\delta<P_j, \xi^l>=0$. Thus we have $h^l=\cH
\xi^l\in \lieh$. Moreover, $L_{h^l}\cH=0$ from Lemma \ref{lem2}.  Since
$$0=L_{h^l}\Re=(L_{h^l} \cH) \J+\cH L_{h^l}\J=\cH L_{h^l}\J ,$$
this leads to $ L_{h^l}\J=0$ due to the non-degeneracy of $\cH$. This implies that
\begin{equation}\label{condd}
\begin{array}{l}\sum_{k=1}^n {\tilde \epsilon}_k  (L_{h^l}\gamma_k \otimes
D_x^{-1} \gamma_k+\gamma_k \otimes D_x^{-1} L_{h^l}\gamma_k )=0 .
\end{array}
\end{equation}
It follows from Theorem \ref{apth} in Appendix that $\tilde \epsilon_i
L_{h^l}\gamma_i=\sum_{k=1}^n \gamma_k A_{ki}, $ where the $A_{ki}$ are constant and
$A_{ki}=-A_{ik}$. From assumption \ref{con4}, if the matrix $A\neq 0$, then $h^l$
contains the vector in $\mbox{Span}_{\bbbc}<P_1, \cdots, P_n>$, which cannot be true
since $\xi^0$ is proper with respect to operator $\cH$ and $\J$. Therefore, the matrix
$A$ must be zero and this implies $L_{h^l}\gamma_k=0$. By Lemma \ref{lem3}, we have
$\delta < h^l, \gamma_k>=0$ and thus $\xi^{l+1}=\J h^l\in \Omega^1$. Using the Leibnitz
rule for the Lie derivative and the assumption $L_{P_j}\J=L_{P_j}\cH=0$, we obtain
$L_{P_j} \xi^{l+1}=L_{P_j}(\J \cH \xi^l)=0$.

Finally, we prove that $\{f^i, f^{i_1}\}=0$ if $\xi^i=\delta f^i$ in the same way as
we did for Theorem \ref{th1}. We have $\{f^i, f^{i_1}\}=<\xi^i, \cH
\xi^{i_1}>=<(\J\cH)^i \xi^{0},\cH (\J\cH)^{i_1}\xi^0> =<\xi^{0}, \Re^{i_1+i}h^0> $
and
\begin{equation*}
\begin{array}{l}\quad \delta <\xi^0,\ \Re^{i_1+i} h^0>=L_{\Re^{i_1+i} h^0}\xi^0=
\delta <\J \cH \xi^0,\ \Re^{i_1+i-1} h^0> =L_{\Re^{i_1+i-1} h^0} (\J \cH \xi^0)\\
=(L_{\Re^{i_1+i-1} h^0} (\J \cH)) \xi^0+\J \cH L_{\Re^{i_1+i-1} h^0} (\xi^0)=\J \cH
L_{\Re^{i_1+i-1} h^0} (\xi^0). \end{array}
\end{equation*}
By induction, we obtain $L_{\Re^{i_1+i} h^0}\xi^0=\delta <\xi^0,\ \Re^{i_1+i} h^0>=0$
implying $\left\{f^i, \ f^{i_1}\right\}=0$ and this completes the proof. \hfill
$\blacksquare$

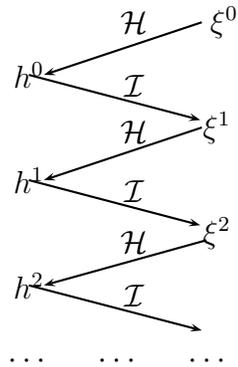
\begin{figure}[ht]
\begin{eqnarray*}
\begin{array}{c}
\begin{pspicture}(0,0)(5,5)
\rput(2.6,4.5){$\xi^0$} \rput(0,3.8){$h^0$}\rput(1.4,4.5){$\cH$}
\psline{->}(2.3,4.5)(0.2,3.8) \rput(2.5,3.1){$\xi^1$}
\rput(0,2.4){$h^1$}\rput(1.4,3.05){$\cH$} \psline{->}(2.3,3.1)(0.2,2.4)
\rput(2.5,1.7){$\xi^2$} \rput(0,1.0){$h^2$}\rput(1.4,1.58){$\cH$}
\psline{->}(2.36,1.6)(0.2,1.0) \psline{->}(0,3.8)(2.28,3.2)\rput(1.4,3.66){$\J$}
\psline{->}(0,2.4)(2.28,1.8)\rput(1.4,2.26){$\J$}\psline{->}(0,1.0)(2.3,0.4)
\rput(1.4,0.86){$\J$}\rput(0,0){$\cdots$}\rput(1.2,0){$\cdots$}\rput(2.4,0){$\cdots$}
\end{pspicture}
\end{array}
\end{eqnarray*}
\caption{Lenard scheme for compatible Hamiltonian and symplectic\newline operators
when the generating Nijenhuis operator is not weakly nonlocal}\label{fig2}
\end{figure}

This theorem gives rise to the Lenard scheme as shown in Figure \ref{fig2}.

Before we apply it to concrete examples, we make a few remarks.
\begin{Rem}\label{rem1}
The assumption \ref{con4} was inspired by Lemma \ref{lem5}. From the proof of the
theorem, the purpose of Definition \ref{defp} and this assumption is to enable us to
draw the conclusion $L_h^{l+1}\xi_k=0$ from identity (\ref{condd}). For concrete
examples, it is possible we do not require such strong assumptions. If $\gamma_k$ is
polynomial and we restrict $Q$ to be polynomial,  we only need to check that $h^l$
does not contain linear terms in dependent variables since only such vector fields
preserve the order and the degree of $\gamma_k$.
\end{Rem}
\begin{Rem}\label{rem2}
In application to nonlinear evolution equations, the $\gamma_k$ in Theorem \ref{th2} are
good candidates for $\xi^0$. If one of them, say $\gamma_n$ without loss of
generality, is indeed the starting point, we only need to check the assumptions
\ref{con3} and \ref{con4} in Theorem \ref{th2} for $\gamma_k,\ k=1,\cdots, n-1$.
\end{Rem}

\begin{Pro} Starting from $\xi^0=\frac{\sqrt 3}{3}\gamma_3$ the hierarchy of
commuting local symmetries and conservation laws for system (\ref{SS}) (cf.
Definition \ref{def}) can be generated using the Lenard scheme as shown in Figure
\ref{fig2}  .
\end{Pro}
{\bf Proof}. To prove the statement, we check the conditions in Theorem \ref{th2}.
The operator $\cH$ is non-degenerate since the determinant of the coefficient
matrix of $D_x$ is non-zero. Following Remark \ref{rem2}, we only consider $\gamma_1$
and $\gamma_2$, which are obviously linear independent over $\mathbb{C}$. Secondly,
by direct calculation we have $L_{P_1} \cH = L_{P_1} \J =0$.  Finally, as we remarked
in Remark \ref{rem1} we only need to check for all $l$ that the vectors $h^l$ do not contain
any terms linear in $u$ and $v$, which is true for the given operators and $\xi^0$
(although assumption \ref{con4} has been proved in Lemma \ref{lem5}).

Now we check the conditions on $\xi^0$. Notice that $\xi^0$ is closed and $L_{P_1}
\xi^0 =0$. Moreover,
\begin{eqnarray*}
\cH \xi^0=\left(\!\!\begin{array}{c}u_x \\ v_x\end{array}\!\!\right); \qquad \J\cH
\xi^0=\left(\!\!\begin{array}{c}v_{xx}+4 u v^2 \\ u_{xx} +4 u^2 v
\end{array}\!\!\right)=\delta \int (-u_x v_x+2 u^2 v^2) .
\end{eqnarray*}
So all the conditions of Theorem \ref{th2} are satisfied and thus we proved the
statement. \hfill $\blacksquare$
\begin{Ex}\label{eg2} Consider the vector modified KdV equation \cite{mr95k:35154}
\begin{equation}\label{mkdv} V_t = V_{xxx} + 3 <V,V>V_x + 3 <V,V_x> V,
\end{equation}
where the dimension of vector $V$ is $N$. Let $J_{ij}$ and  $S_{ij}$ be $N\times N$
matrices. We represent its Hamiltonian operator and sympletic operator \cite{anco6}
as follows:
\begin{eqnarray*}
&&\cH=D_x+\sum_{1\leq i<j\leq N} (J_{ij}V) \otimes D_x^{-1} (J_{ij}V);
\quad (J_{ij})_{kl}=\delta_i^k\delta_j^l-\delta_i^l\delta_j^k\\
&&\J=D_x +2 V \otimes D_x^{-1} V +\!\!\!\!\!\sum_{1\leq i\leq j\leq
N}\!\!\!\frac{1}{1+\delta_i^j} (S_{ij} V) \otimes D_x^{-1} (S_{ij} V);\quad
(S_{ij})_{kl}=\delta_i^k\delta_j^l+\delta_i^l\delta_j^k.
\end{eqnarray*}
\end{Ex}
\begin{Pro}\label{pkdv} Starting from $\xi^0=V$ the hierarchy of commuting local symmetries
and conservation laws for equation (\ref{mkdv}) can be generated using the Lenard
scheme as shown in Figure \ref{fig2}.
\end{Pro}
We first check assumption \ref{con2} of Theorem \ref{th2} in the following lemma.
\begin{Lem}\label{TmpL} The Lie derivatives of $\cH$ and $\J$ as defined in Example \ref{eg2}
along the vector $J_{ij} V$ vanish, that is, $L_{J_{ij} V} \cH=L_{J_{ij} V} \J=0.  $
\end{Lem}
{\bf Proof}. According to Definition \ref{def}, we have
\begin{eqnarray*}
&&L_{J_{ij} V} \cH= D_{\cH}[J_{ij} V] -J_{ij} \cH+\cH J_{ij}\\
&&\quad =\sum_{1\leq k<l\leq N}\left\{ ([J_{kl}, \  J_{ij}] V) \otimes D_x^{-1}
(J_{kl}V) + (J_{kl}  V) \otimes D_x^{-1} ([J_{kl},\ J_{ij}] V)\right\};
\\&&L_{J_{ij} V} \J= D_{\J}[J_{ij} V] -J_{ij} \J+\J J_{ij}\\
&&\quad =\sum_{1\leq k\leq l\leq N}\left\{ ([S_{kl}, \  J_{ij}] V) \otimes D_x^{-1}
(J_{kl}V) + (J_{kl}  V) \otimes D_x^{-1} ([S_{kl},\ J_{ij}] V)\right\}.
\end{eqnarray*}
Using the fact that
\begin{eqnarray*}&&[J_{kl}, \  J_{ij}]=J_{kj}\delta_{l}^i+J_{ik} \delta_{l}^j
+J_{jl} \delta_{k}^i+J_{li}\delta_{k}^j;   \\
&&[S_{kl}, \ J_{ij}]=S_{kj}\delta_{l}^i-S_{ik} \delta_{l}^j +S_{jl}
\delta_{k}^i-S_{li}\delta_{k}^j;
\end{eqnarray*}
we can prove that $L_{J_{ij} V} \cH=L_{J_{ij} V} \J=0$. \hfill $\blacksquare$

{\bf Proof of Proposition \ref{pkdv}}. We check the conditions in Theorem \ref{th2}
one by one. First the operator $\cH$ is non-degenerate since the determinant of the
coefficient matrix of $D_x$ is non-zero. For assumption \ref{con3}, following Remark
\ref{rem2}, we only consider $S_{ij} V$, $1\leq i\leq j\leq$, which are linear
independent over $\mathbb{C}$. Assumption \ref{con2} is proved in Lemma \ref{TmpL}.
Finally, according to Remark \ref{rem1}, instead of checking assumption
\ref{con4}, we only need to check for all $l$ that the vectors $h^l$ do not contain any terms
linear in $u$ and $v$, which is true for the given operators and $\xi^0$.

Now we check the conditions on $\xi^0$. Notice that $\xi^0$ is closed and $L_{J_{ij}
V} \xi^0 =0$. Moreover,
\begin{eqnarray*}
\cH \xi^0=V_x; \qquad \J\cH \xi^0=V_{xx}+2 <V, V> V =\delta \int -\frac{1}{2} (<V_x,
V_x>-<V, V>^2) .
\end{eqnarray*}
So following the proof of Theorem \ref{th2} we obtain the results in the statement.
\hfill $\blacksquare$

The following lemma show how to check assumption \ref{con4} for Example \ref{eg2}
although it is not necessary since the objects we considered are differential
polynomials.

\begin{Lem}\label{lem6}
Let $J_{ij} V$ and $S_{ij} V $ be defined as (\ref{eg2}) in Example \ref{eg1}. For
any $Q\in \lieh$ and anti-symmetric constant matrix $A$, if
\begin{equation}\label{rela}
\begin{array}{c}\frac{1}{1+\delta_i^j} L_Q (S_{ij} V) =\sum_{1\leq k\leq l\leq N}
A_{klij} (S_{kl}V)\end{array},
\end{equation}
where $A_{klij}=-A_{ijkl}$,
$A_{klij}=A_{lkij}$ and $A_{klij}=A_{klji}$, then $Q\equiv \mbox{Span}_{\bbbc}\!\!<
\!\!J_{ij},1\!\!\leq i< j\leq N \!\!>$ with respect to $\{S_{ij}, 1\leq i\leq j\leq
N\}$.
\end{Lem}
{\bf Proof}. When $i=j$, we know $\frac{1}{2}L_{Q} (S_{ii} V)=\sum_{1\leq k\leq l\leq
n} A_{kl ii} (S_{kl}V)$. From Definition \ref{def0}, it follows that the $k$-th
component of $Q$ is equivalent to $\sum_{j=1}^N A_{kjkk} V^{(j)} $, where
$A_{kjkk}=A_{jkkk}$ and $V^{(j)}$ is the $j$-th component of the vector $V$. For such
$Q$ and $i<j$, we compute
\begin{eqnarray*}
\begin{array}{c}L_Q (S_{ij} V)=\sum_{l=1}^N (A_{ilii} S_{lj}V+A_{jljj}S_{li}
V)\end{array}.
\end{eqnarray*}
Comparing to the coefficient of $S_{ii}V$ in (\ref{rela}), we obtain that $A_{iiij} =
A_{jijj}=-A_{jjij}$. This leads to $Q\equiv \sum_{1\leq i<j\leq n} A_{ijii} J_{ij}V $
with respect to $\{S_{ij}, 1\leq i\leq j\leq N\}$. \hfill $\blacksquare$

\section{Construction of recursion operators}\label{sec4}
In this section, we construct a recursion operator of system (\ref{sys}) from its Lax
representation. In general, it is not easy to construct a recursion operator for a
given integrable equation although the explicit formula is given, cf. Definition
\ref{def}. The difficulty lies in how to determine the starting terms of $\Re$, i.e.,
the order of the operator, and how to construct its nonlocal terms. Many papers are
devoted to this subject, see \cite{mr88g:58080,gu98,mr1974732}. If the Lax
representation of the equation is known, there is an amazingly simple approach to
construct a recursion operator proposed in \cite{mr2001h:37146}. The idea in
\cite{mr2001h:37146} can be developed for the Lax pairs that are invariant under the
reduction groups. The general setting up and results will be published later. Here we
only treat system (\ref{sys}).

\subsection{Integrability of evolution equations}\label{sec3}
Before we proceed, we first give the basic definitions for symmetries, cosymmetries
and recursion operators, etc. for evolution equations \cite{mr94j:58081,mr94g:58260}
in the context of the variational complex described in Section \ref{sec2}. Meanwhile
we fix the notation.

To each element $K\in \lieh$, we can associate an evolution equation of the form
\begin{eqnarray}\label{eq}
u_t=K.
\end{eqnarray}
Strictly speaking, this association is not as innocent as it looks, since one
associates to the evolution equation the derivation
$$
\ddt+ \sum_{k=0}^\infty D_x^k K \pderiv{}{u_k}.
$$
As long as objects concerned are time-independent as in this paper, one does not see
the difference, but in the time dependent case one really has to treat the equation
and its symmetries as living in different spaces, cf. \cite{wang98} for details.

\begin{Def}\label{def}
Given an evolution equation (\ref{eq}), when Lie derivatives of the following vanish
along $K\in \lieh$ we call: $\int\!\! g\in\A'$ a conserved density; $h\in\lieh$  a
symmetry; $\xi\in\Omega^1$ a cosymmetry; $\Re: \lieh \rightarrow \lieh$ a recursion
operator; a Hamiltonian operator $\cH: \Omega^1 \rightarrow \lieh$ a Hamiltonian
operator for the equation; a symplectic operator $\J: \lieh \rightarrow \Omega^1 $ a
symplectic operator for the equation.
\end{Def}

Here we use the standard definition of a recursion operator in the literature. We
refer the reader to \cite{mr2002b:37100} for a discussion of the problems with this
definition when symmetries are time-dependent.

From the above definitions, we can show that if $\int\!\! f$ is a conserved density
of the equation, then $\delta(\int\!\! f)$ is its cosymmetry. Moreover, if $\cH$ is a
Hamiltonian operator and $\J$ is a symplectic operator of a given equation, then $\cH
\J$ is a recursion operator. Operator $\cH$ maps cosymmetries to symmetries while
$\J$ maps symmetries to cosymmetries.

We say that the evolution equation (\ref{eq}) is a Hamiltonian system if for a
(pseudo-differential) Hamiltonian operator $\cH$, there exists a functional $\int\!\!
f$, called the Hamiltonian, such that $\cH\ \delta(\!\int\!\! f) $ is a symmetry of the
equation. Additionally, if for a (pseudo-differential) symplectic operator $\J$,
which is compatible with $\cH$, there exists a functional $\int\! g$ such that
\begin{eqnarray*}
 \J u_t= \J K= \delta(\!\int\!\! g)\ ,
\end{eqnarray*}
we say that the evolutionary equation is a (generalised) bi-Hamiltonian system.

The Sawada-Kotera equation (\ref{sk}) is a bi-Hamiltonian system since we have
$$u_t=\cH \delta\int\!\!(-\frac{u_x^2}{2}+\frac{u^3}{6} )$$ and
$$\J u_t=\delta\int\!\!\left(\frac{1}{2}u_{4x}^2-\frac{7}{2}u u_{3x}^2+\frac{8}{3}u_{2x}^3
+8 u^2 u_{2x}^2-\frac{17}{6}u_x^4-\frac{25}{3}u^3 u_x^2+\frac{2}{9}u^6  \right). $$

\subsection{Construction of a recursion operator}\label{sec42}
Consider a matrix operator of the form
\begin{eqnarray}\label{lax}
L(\la)=D_x+\la^{-1} V \dl-\la \dl^{-1} V,
\end{eqnarray}
where $\la$ is the spectral parameter and $\dl$ is a $3\times 3$ matrix satisfying
$$\dl =\left(\begin{array}{lcr}0&1&0\\0&0&1\\1&0&0\end{array}\right)\qquad
\mbox{and}\qquad V=\left(\begin{array}{lcr}v_1&0&0\\0&v_2&0\\0&0&v_3\end{array}
\right) = \left(\begin{array}{lcr}e^{\phi_1}&0&0\\0& e^{\phi_2}&0\\0&0&e^{\phi_3}
\end{array} \right).$$
Here $\dl$ acts as a shift operator. Clearly  we have $\dl^3=I$ and
$\dl^T=\dl^{-1}=\dl^2$.
\begin{Not} From now on, we often write $v_i=e^{\phi_i}$, $i=1,..,3$, where $\sum_{i=1}^3
\phi_i=0$.
\end{Not}

The operator $L(\la)$ is invariant under the following two transformations
$$
s: L(\la)\mapsto S^{-1} L(\sigma \la) S \qquad \mbox{and} \qquad r: L(\la)\mapsto -
L^{\star}(\frac{1}{\la}),
$$
where S is a diagonal $3\times 3$ matrix given by $S_{ii}=\sigma^{i}$ and
$\sigma=e^{2\pi i/3}$. These two transformations satisfy
$$s^2 = r^2 = id,\quad rsr = s^{-1} $$
and therefore generate the dihedral group $\mathbb D_3$. The reduction groups of Lax
pairs have been studied in \cite{mik81,LM04,LM05}.

Assume that $\sum_{j=1}^3 \phi_j=0$. Consider the zero curvature equation
\begin{eqnarray}\label{zero}
\begin{array}{c}L(\la)_t=[ L(\la),\ A(\la)],
\end{array}
\end{eqnarray}
where $A(\la)=\lambda^{-1} \mathbf{b}V \dl - \lambda \dl^{-1} V \mathbf{b} + 3
\lambda^{-2} (V \Delta)^2 - 3 \lambda^2 (\Delta^{-1} V)^2$ and $\mathbf{b}$ is a
diagonal $3\times 3$ matrix with entries $\mathbf{b}_{ii}=\phi_{i+1,x}-\phi_{i+2,x} $
under the convention $\phi_{i}=\phi_{i\!\!\mod\!3}$ for $i>3$. It gives us a
3-component system
\begin{eqnarray}\label{3zero}
\phi_{i,t}=\phi_{i+1,xx}-\phi_{i+2,xx}+(\phi_{i+1,x}-\phi_{i+2,x}) \phi_{i,x}+3 e^{2
\phi_{i+1}}-3 e^{2 \phi_{i+2}}, \qquad i=1,2,3.
\end{eqnarray}
Substituting the constraint $\sum_{j=1}^3 \phi_j=0$ into (\ref{3zero}), we obtain
system (\ref{sys}).

Notice that system (\ref{3zero}) is homogeneous if we assign the weights of
$v_i=e^{\phi_i}$ as $1$ and the weights of $\phi_i$ as zero. We can also consider
system (\ref{sys}) to be homogeneous under the same weights since we derive it from
homogeneous system (\ref{3zero}). This homogeneity enables us directly apply the
results in \cite{mr1974732}.

The operator $A(\la)$ in (\ref{zero}) is also invariant under the transformations $s$
and $r$. In the commutator of the operators $L(\la)$ and $A(\la)$, the coefficients
of positive powers of $\la $ are transposes of the negative powers and thus give no
extra information. The constant term, i.e., the coefficient of $\lambda^0$, is
trivially satisfied. This is true in general.

For given $L(\la)$, we can build up a hierarchy of nonlinear systems by choosing
different operators $A(\la)$ starting with $ \lambda^{-n} (V \dl)^n$. It is easy to
check $(V \dl)^3=I$ when $\sum_{j=1}^3 \phi_j=0$. This implies that system
(\ref{sys}) has no symmetries of order $3n$.

The idea to construct a recursion operator directly from a Lax representation is to
relate the different operators $A(\la)$ using ansatz ${\bar A}(\la)={\cal P} A(\la)
+R$ and then to find the relation between two flows corresponding to $\bar A(\la)$
and $A(\la)$. Here ${\cal P}$ commutes with $L(\la)$ and $R$ is the reminder
\cite{mr2001h:37146}.

Since the operator $L(\la)$ given by (\ref{lax}) is invariant, we require that the
ansatz ${\cal P} A(\la) +R$ is also invariant. We take ${\cal P}=\la^3+\la^{-3}$,
which is a primitive automorphic function of the group $\mathbb{D}_3$ and $R$ is of
the form
\begin{eqnarray}\label{ansa}
R=\sum_{j=1}^3 \left(\la^{-j} d_j (V\dl)^j-\la^j(\dl^{-1}V)^j d_j\right) \quad
\mbox{where} \quad
d_j=\left(\begin{array}{lcr}d_{j,1}&0&0\\0&d_{j,2}&0\\0&0&d_{j,3}\end{array}\right).
\end{eqnarray}
This leads to
\begin{eqnarray}\label{Lax}
\begin{array}{l}
L(\la)_{\tau}=[{\cal P} A(\la) +R, L(\la)]={\cal P} L(\la)_t +[R, \ L(\la)].
\end{array}
\end{eqnarray}
Substituting the ansatz (\ref{ansa}) into (\ref{Lax}) and collecting the coefficient
of negative powers of $\la$, we obtain
\begin{eqnarray}\label{rel}
\begin{array}{ll}
\la^{-4}:    & V_t \dl- V \dl d_{3} + d_{3} V\dl=0;\\
\la^{-3}:   &  d_2-D_x d_3- V \dl d_{2} (V\dl)^{2}=0;\\
\la^{-2}:   & - \dl^{-1} V_t - D_x(d_{2} (V\dl)^{2})- V \dl d_{1} (V\dl)  + \dl^{-1}
V d_{3} + d_{1} (V\dl)^{2}- d_{3} \dl^{-1} V=0;\\
\la^{-1}:    & V_{\tau} \dl= -D_x(d_1 V\dl)+\dl^{-1} V d_{2} (V\dl)^{2}-d_{2}
(V\dl)^{2} \dl^{-1} V.
\end{array}
\end{eqnarray}
We introduce the notation
$$\phi=\left(\begin{array}{lcr}\phi_{1}&0&0\\0&\phi_{2}&0\\0&0&\phi_{3}\end{array}\right)
$$
and define $ \dl^i \phi=\phi^{(i)} \dl$, where $i=1,2$. Using this notation, we have
$V_t=V \phi_t$ and $D_x (V\dl)^j=\sum_{i=0}^{j-1} \phi_x^{(i)} (V\dl)^j .$ Now
formula (\ref{rel}) can be simplified as follows
\begin{eqnarray}
&& \phi_t -  d_{3}^{(1)} + d_{3} =0;\label{d3}\\
&&  D_x d_3+ d_{2}^{(1)}-d_{2}=0;\label{d2}\\
&& -2 \dl^{-1} \phi_t V^2 \dl - D_x(d_{2})- d_{2}\phi_x
-d_2 \phi_x^{(1)} -d_{1}^{(1)}+d_{1} =0;\label{d1}\\
&&\phi_{\tau}= -D_x(d_1) -d_1 \phi_x+\dl^{-1} d_{2} V^2\dl-d_{2} {V^2}^{(1)}.
\label{d0}
\end{eqnarray}
Under the assumption $\sum_{j=1}^3 \phi_j=0$, we can solve system (\ref{d3}). The
unique solution for traceless matrix $d_3$ is
\begin{eqnarray*}
d_3=-\frac{1}{3} (2\phi_t+\phi_t^{(1)})=\frac{1}{3} (\phi_t^{(2)}-\phi_t).
\end{eqnarray*}
Since $d_3$ is traceless, system (\ref{d2}) is consistent. Its general solution is
\begin{eqnarray*}
d_{2}&=&\frac{1}{3}(2 D_x d_3+ D_x d_3^{(1)}) +c_{2} I =\frac{1}{3}
\phi_{xt}^{(2)}+c_{2} I ,
\end{eqnarray*}
where $c_{2}$ is a constant. In order to solve system (\ref{d1}), the trace of its
both sides should be equal, that is,
\begin{eqnarray*}
0&=& \mbox{Tr}(-2 \dl^{-1} \phi_t V^2 \dl - D_x(d_{2})- d_{2}\phi_x
-d_2 \phi_x^{(1)})\\
&=&-D_t(e^{2 \phi_1}+e^{2 \phi_2}+e^{2 \phi_3})-3 D_x c_2+\frac{1}{6} D_t
(\phi_{1,x}^2 + \phi_{2,x}^2+\phi_{3,x}^2),
\end{eqnarray*}
where $\mbox{Tr}$ denotes the trace of a matrix. So we can take
\begin{eqnarray*}
\begin{array}{l}
c_2 =-\frac{1}{3}\sum_{j=1}^3 D_x^{-1}(v_j^2-\frac{1}{6} \phi_{j,x}^2)_t .
\end{array}
\end{eqnarray*}
We now substitute $d_2$ into (\ref{d1}). It follows
\begin{eqnarray*}
\begin{array}{l}
-2 \phi_t^{(2)} {V^2}^{(2)}  -\frac{1}{3} \phi_{xxt}^{(2)}-(D_x c_2)
I+\frac{1}{3}\phi_x^{(2)} \phi_{xt}^{(2)}+c_2 \phi_x^{(2)} -d_{1}^{(1)}+ d_{1} =0.
\end{array}
\end{eqnarray*}
Solving for $d_{1}$, we obtain
\begin{eqnarray*}
\begin{array}{l}
d_1=\frac{1}{3}\left(2 \phi_t^{(2)} {V^2}^{(2)}  +\frac{1}{3}
\phi_{xxt}^{(2)}-\frac{1}{3}\phi_x^{(2)} \phi_{xt}^{(2)}-c_2 \phi_x^{(2)}\right.\\
\quad \left. -2 \phi_t^{(1)} {V^2}^{(1)}  -\frac{1}{3}
\phi_{xxt}^{(1)}+\frac{1}{3}\phi_x^{(1)} \phi_{xt}^{(1)}+c_2 \phi_x^{(1)} \right)
+c_1 I,
\end{array}
\end{eqnarray*}
where $c_1$ is a constant. Again due to the consistence of system (\ref{d0}), $c_1$
satisfies
\begin{eqnarray*}
0&=& -3D_x c_1- \mbox{Tr} (d_1 \phi_x- d_{2}^{(2)} {V^2}^{(2)}+d_{2} {V^2}^{(1)}) .
\end{eqnarray*}
Using the solutions of $d_2$ and $d_1$, we can write it as
\begin{eqnarray*}
3D_x c_1=\frac{1}{9}\left(\begin{array}{ccc} \alpha_1,& \alpha_2, & \alpha_3
\end{array}\right) \left(\begin{array}{c}\phi_{1,t}\\\phi_{2,t}\\
\phi_{3,t}\end{array}\right),
\end{eqnarray*}
where $\alpha_i= (\phi_{i+2,x}-\phi_{i+1,x})(D_x^2-\phi_{i,x} D_x +6 v_{i}^2)+3
(v_{i+1}^2-v_{i+2}^2) D_x .$ Now we have determined $d_2$ and $d_1$. Substituting
them into (\ref{d0}), we have
\begin{eqnarray*}
\begin{array}{ll}
3 \phi_{i,\tau}=D_x \left(-2 \phi_{i+2,t} v_{i+2}^2  -\frac{1}{3}
\phi_{i+2,xxt}+\frac{1}{3}\phi_{i+2,x} \phi_{i+2,xt}-c_2 \phi_{i+2,x}\right.\\
\qquad\quad \left. +2 \phi_{i+1,t} v_{i+1}^2  +\frac{1}{3}
\phi_{i+1,xxt}-\frac{1}{3}\phi_{i+1,x} \phi_{i+1,xt}+c_2 \phi_{i+1,x} +3 c_1\right)\\
\qquad\quad + \phi_{i,x} \left(-2 \phi_{i+2,t} v_{i+2}^2 -\frac{1}{3}
\phi_{i+2,xxt}+\frac{1}{3}\phi_{i+2,x} \phi_{i+2,xt}-c_2 \phi_{i+2,x}\right.\\
\qquad\quad\left. +2 \phi_{i+1,t} v_{i+1}^2  +\frac{1}{3}
\phi_{i+1,xxt}-\frac{1}{3}\phi_{i+1,x} \phi_{i+1,xt}+c_2 \phi_{i+1,x} +3 c_1\right)\\
\qquad\quad+ (\phi_{i+1,xt}v_{i+2}^2-\phi_{i+2,xt}v_{i+1}^2)+ 3 c_2
(v_{i+1}^2-v_{i+2}^2)\\
\qquad=D_x \left(-2 \phi_{i+2,t} v_{i+2}^2  -\frac{1}{3}
\phi_{i+2,xxt}+\frac{1}{3}\phi_{i+2,x} \phi_{i+2,xt} +2 \phi_{i+1,t} v_{i+1}^2
+\frac{1}{3}
\phi_{i+1,xxt}-\frac{1}{3}\phi_{i+1,x} \phi_{i+1,xt}\right)\\
\qquad\quad+ \frac{1}{3}\sum_{j=1}^3(2 v_j^2 \phi_{j,t}-\frac{1}{3} \phi_{j,x}
\phi_{j,xt})
(\phi_{i+1,x}-\phi_{i+2,x})+(\phi_{i+1,xt}v_{i+2}^2-\phi_{i+2,xt}v_{i+1}^2) \\
\qquad\quad-\frac{1}{9}\sum_{j=1}^3((\phi_{j+2,x}-\phi_{j+1,x})(6 v_j^2
\phi_{j,t}+\phi_{j,xxt}-\phi_{j,x}\phi_{j,xt})+
3(v_{j+1}^2-v_{j+2}^2)\phi_{j,xt})\\
\qquad\quad+ \phi_{i,x} \left(-2 \phi_{i+2,t} v_{i+2}^2 -\frac{1}{3}
\phi_{i+2,xxt}+\frac{1}{3}\phi_{i+2,x} \phi_{i+2,xt}  +2 \phi_{i+1,t} v_{i+1}^2
+\frac{1}{3}
\phi_{i+1,xxt}-\frac{1}{3}\phi_{i+1,x} \phi_{i+1,xt} \right)\\
\qquad\quad-\frac{1}{9}\left(3(v_{i+1}^2-v_{i+2}^2)+(\phi_{i+1,xx}-\phi_{i+2,xx})
+\phi_{i,x}(\phi_{i+1,x}-\phi_{i+2,x})\right)\sum_{j=1}^3 \phi_{j,x} \phi_{j,t}\\
\qquad\quad-\frac{1}{9}\phi_{i,x}\sum_{j=1}^3((\phi_{j+2,x}-\phi_{j+1,x})
(\phi_{j,xt}-\phi_{j,x}\phi_{j,t})+
3(v_{j+1}^2-v_{j+2}^2)\phi_{j,t}-(\phi_{j+2,xx}-\phi_{j+1,xx})\phi_{j,t})\\
\qquad\quad+\frac{1}{3}\left(3(v_{i+1}^2-v_{i+2}^2)+(\phi_{i+1,xx}-\phi_{i+2,xx})
+\phi_{i,x}(\phi_{i+1,x}-\phi_{i+2,x})\right) D_x^{-1}\sum_{j=1}^3(2 v_j^2
+\frac{1}{3}
\phi_{j,xx}) \phi_{j,t}\\
\qquad\quad-\frac{1}{9}\phi_{i,x} D_x^{-1}\sum_{j=1}^3\left(
(\phi_{j+2,x}-\phi_{j+1,x})(6
v_j^2+\phi_{j,xx})+ (\phi_{j+2,xxx}-\phi_{j+1,xxx})\right.\\
\qquad\qquad\quad\left.+\phi_{j,x}(\phi_{j+2,xx}-\phi_{j+1,xx})
-6(v_{j+1}^2\phi_{j+1,x}-v_{j+2}^2\phi_{j+2,x})\right) \phi_{j,t}\ .
\end{array}
\end{eqnarray*}
We substitute $\phi_3=-\phi_1-\phi_2$ into the above expression. This leads to a
recursion operator $\Re$ of system (\ref{sys}) mapping the flow
$\left(\begin{array}{lc} \phi_{1,t}, & \phi_{2,t}
\end{array} \right)^T$ to the flow $\left(\begin{array}{lc} \phi_{1,\tau}, &
\phi_{2,\tau } \end{array} \right)^T$, that is,
$$\left(\begin{array}{l} \phi_{1,\tau}\\\phi_{2,\tau } \end{array} \right)
=\Re\left(\begin{array}{l} \phi_{1,t}\\\phi_{2,t} \end{array} \right)
=\left(\begin{array}{lr} \Re_{11} & \Re_{12}\\\Re_{21} & \Re_{22} \end{array} \right)
\left(\begin{array}{l} \phi_{1,t}\\\phi_{2,t} \end{array} \right).
$$
\begin{The}\label{ReS}
System (\ref{sys}) possesses a recursion operator of order $3$ with entries
\begin{eqnarray*}
\begin{array}{ll}
\Re_{11}(\phi_1,\phi_2) =\frac{1}{3} D_x^3 +\frac{2}{3}( \phi_{1,x}+\phi_{2,x})
D_x^2\\\qquad + \frac{1}{9}(3 \phi_{1,xx}- \phi_{1,x} \phi_{2,x}+3\phi_{2,xx}-
\phi_{1,x}^2-\phi_{2,x}^2+21 v_3^2+3 v_2^2 +3 v_1^2)  D_x
\\
\qquad- \frac{1}{9}(7 \phi_{1,x} \phi_{2,xx}+ \phi_{1,xx} \phi_{2,x} +2 \phi_{1,x}
\phi_{1,xx} +2 \phi_{2,x} \phi_{2,xx}+4
\phi_{1,x}^3 +7 \phi_{1,x}^2 \phi_{2,x} +\phi_{1,x} \phi_{2,x}^2)\\
\qquad +\frac{1}{3}(5  v_1^2  -4  v_2^2  -7 v_3^2) \phi_{1,x} +\frac{1}{3}(8
v_1^2-v_2^2-13  v_3^2) \phi_{2,x}
\\\qquad -\frac{1}{9} \phi_{1,x} D_x^{-1} s_{2,\phi_1} -\frac{1}{9} \phi_{1,t} D_x^{-1}
s_{1,\phi_1}
;\\
\Re_{12}(\phi_1,\phi_2)=\frac{2}{3} D_x^3 +\frac{2}{3} \phi_{1,x}D_x^2+\frac{1}{9} (3
\phi_{1,xx}-2 \phi_{1,x}^2-2 \phi_{1,x} \phi_{2,x} -2 \phi_{2,x}^2 +6 v_1^2+24 v_2^2
+24 v_3^2) D_x
\\\qquad+\frac{2}{9} (\phi_{1,x} \phi_{1,xx}-\phi_{1,x} \phi_{2,xx}-\phi_{2,x} \phi_{1,xx}
-2 \phi_{2,x} \phi_{2,xx} -\phi_{1,x}^3 -\phi_{1,x}^2 \phi_{2,x} -\phi_{1,x}
\phi_{2,x}^2 )
\\\qquad+\frac{2}{3} (v_1^2+v_2^2-5 v_3^2) \phi_{1,x}+4 (v_2^2-v_3^2) \phi_{2,x}
-\frac{1}{9} \phi_{1,x} D_x^{-1} s_{2,\phi_2} -\frac{1}{9} \phi_{1,t} D_x^{-1}
s_{1,\phi_2} ;
\end{array}
\end{eqnarray*}
and $\Re_{21}(\phi_1,\phi_2)=-\Re_{12}(\phi_2,\phi_1)$;
$\Re_{22}(\phi_1,\phi_2)=-\Re_{11}(\phi_2,\phi_1)$. Here
\begin{eqnarray}
&&\left\{\begin{array}{ll} s_{1,\phi_1}=-2 \phi_{1,xx}-\phi_{2,xx}
-6 e^{2 \phi_{1}}+6 e^{-2 \phi_1-2\phi_2}\\
s_{1,\phi_2}=-\phi_{1,xx}-2\phi_{2,xx}  -6 e^{2 \phi_{2}} +6 e^{-2 \phi_1-2\phi_2}
\end{array}\right.\label{cos1}\\
&&\left\{\begin{array}{ll} s_{2,\phi_1}=-3 \phi_{2,xxx} -4  \phi_{1,x} \phi_{1,xx}+ 2
\phi_{2,x} \phi_{2,xx}-2 \phi_{1,xx} \phi_{2,x}-2 \phi_{1,x} \phi_{2,xx} \\
\qquad \quad -12 e^{2 \phi_{1}} \phi_{2,x} -12 e^{2 \phi_{2}} \phi_{2,x}
-12 e^{-2 \phi_1-2\phi_2} \phi_{2,x}\\
s_{2,\phi_2}=3 \phi_{1,xxx} -2  \phi_{1,x} \phi_{1,xx}+ 4
\phi_{2,x} \phi_{2,xx}+2 \phi_{1,xx} \phi_{2,x}+2 \phi_{1,x} \phi_{2,xx} \\
\qquad \quad +12 e^{2 \phi_{1}} \phi_{1,x} +12 e^{2 \phi_{2}} \phi_{1,x} +12 e^{-2
\phi_1-2\phi_2} \phi_{1,x}
\end{array}\right.\label{cos2}
\end{eqnarray}
and $\phi_{1,t}$ , $\phi_{2,t}$ is the system itself.
\end{The}
In Theorem \ref{ReS}, (\ref{cos1}) and (\ref{cos2}) are two co-symmetries of system
(\ref{sys}). They are variational derivatives of the following two conservation laws:
\begin{eqnarray*}
\begin{array}{l}  H_1=\phi_{1,x}^2+\phi_{1,x} \phi_{2,x}+\phi_{2,x}^2-3 (v_1^2+v_2^2+v_3^2);\\
H_2=3 \phi_{1,x} \phi_{2,xx}+\phi_{1,x}^2 \phi_{2,x}-\phi_{1,x}
\phi_{2,x}^2+\frac{2}{3} \phi_{1,x}^3-\frac{2}{3} \phi_{2,x}^3 +6 v_1^2 \phi_{3,x} +6
v_2^2 \phi_{1,x} +6 v_3^2 \phi_{2,x}.
\end{array}
\end{eqnarray*}
The nonlocal terms of the recursion operator are determined by the symmetries and
co-symmetries of the corresponding orders. Such structures of recursion operators
have been discussed in \cite{mr1974732, serg5}. It has been shown that such recursion
operator gives rise to hierarchies of infinitely many commuting local symmetries if
it is a Njienhuis operator.
\begin{The}\label{RN}
The operator $\Re$ defined in Theorem \ref{ReS} is Nijenhuis.
\end{The}
{\bf Proof}. We need to check that the expression $H:=D_{\Re}[\Re P](Q)-\Re D_{\Re}[P]
(Q)$ is symmetric with respect to two-component vectors $P$
and $Q$. 
We use subindex $i$ to denote $i$-th component. The first component of $H$ will be
written as $H_1$. The calculation is straightforward, but rather complicated. Here we
only pick out the constant terms in $H$, i.e., terms are independent of the dependent
variables $\phi_1$, $\phi_2$ and their $x$-derivatives. We denote these terms by
$H^0$.

For the recursion operator $\Re$, its constant and linear terms are
\begin{eqnarray*}
&&\Re^0=\frac{1}{3} \left(\begin{array}{rr} D_x^3& 2 D_x^3\\
-2 D_x^3& - D_x^3\end{array}\right);\\
&&\Re^1=\frac{1}{3}\left(\begin{array}{cc}( \phi_{1,x}+\phi_{2,x}) D_x^2 + D_x (
\phi_{1,x}+\phi_{2,x})  D_x &  \phi_{1,x}D_x^2+
D_x \phi_{1,x} D_x\\
- \phi_{2,x}D_x^2- D_x \phi_{2,x} D_x& -( \phi_{1,x}+\phi_{2,x}) D_x^2 -D_x (
\phi_{1,x}+\phi_{2,x}) D_x\end{array}\right).
\end{eqnarray*}
So $H^0=D_{\Re^1}[\Re^0 P](Q)-\Re^0 D_{\Re^1}[P] (Q)$. Due to the relations among the
entries of the operator $\Re$, it is easy to see the second component of $H^0$, i.e.,
$(H^0)_2$ is related to its first component as follows: $$ (H^0)_2(P_1, P_2; Q_1,
Q_2)=(H^0)_1(P_2, P_1; Q_2, Q_1).$$ Thus we only require to check whether the first
component is symmetric with respect to $P$ and $Q$ or not. Notice that
\begin{eqnarray*}
\begin{array}{l}
\qquad 9 (H^0)_2(P_1, P_2; Q_1, Q_2)\\
=(P_{2,4x}-P_{1,4x}) Q_{1,2x}+
D_x\left(( P_{2,4x}-P_{1,4x}) Q_{1,x} + (P_{1,4x}+2 P_{2,4x})Q_{2,x}\right)\\
\quad +(P_{1,4x}+2 P_{2,4x})Q_{2,2x}-D_x^3((P_{1,x}-P_{2,x})Q_{1,2x} -( P_{1,x}+2
P_{2,x}) Q_{2,2x})\\
\quad  -D_x^4( ( P_{1,x}+P_{2,x}) Q_{1,x}  +  P_{1,x} Q_{2,x}-2( P_{1,x}+P_{2,x})
Q_{2,x} - 2 P_{2,x} Q_{1,x})\\
=( P_{2,5x}-P_{1,5x}) Q_{1,x} + (P_{1,5x}+2 P_{2,5x})Q_{2,x}
+( P_{2,x}-P_{1,x}) Q_{1,5x} + (P_{1,x}+2 P_{2,x})Q_{2,5x}\\
\quad +3 D_x^2\left((P_{2,2x}-P_{1,2x})Q_{1,2x} +( P_{1,2x}+2 P_{2,2x})
Q_{2,2x}\right)\\\quad -3\left((P_{2,3x}-P_{1,3x})Q_{1,3x} +( P_{1,3x}+2
P_{2,3x}) Q_{2,3x}\right)\\
\quad  -D_x^4\left( ( P_{1,x}+P_{2,x}) Q_{1,x}  +  P_{1,x} Q_{2,x}-2(
P_{1,x}+P_{2,x}) Q_{2,x} - 2 P_{2,x} Q_{1,x}\right).
\end{array}
\end{eqnarray*}
This is symmetric with respect to $P$ and $Q$ and thus we proved the statement. \hfill
$\blacksquare$

This Nijenhuis operator $\Re$ defined in Theorem \ref{ReS} has two seeds: the trivial
symmetry $u_x=\left(\begin{array}{ll} \phi_{1,x}, & \phi_{2,x} \end{array} \right)^T$
and system (\ref{sys}). We can generate the local symmetries of order $3k+1$ and
$3k+2$ for $k\geq 0$ of system (\ref{sys}) by recursively applying the operator $\Re$
on these two seeds. In particular, $\Re (u_x)$ gives us a local symmetry of order $4$
with first component
\begin{eqnarray*}
\begin{array}{l}G(\phi_1,\phi_2)_{\phi_1}=\frac{1}{3}\phi_{1,4x}+\frac{2}{3} \phi_{2,4x}
+\frac{2}{3} \phi_{1,x} \phi_{1,3x}+\frac{2}{3} \phi_{1,x} \phi_{2,3x}+\frac{2}{3}
\phi_{2,x} \phi_{1,3x}+\frac{1}{3}\phi_{1,2x}^2+\frac{2}{3} \phi_{1,2x}
\phi_{2,2x}\\\qquad -\frac{2}{9} (\phi_{1,x}^2+\phi_{1,x} \phi_{2,x}+\phi_{2,x}^2)
(\phi_{1,2x}+2 \phi_{2,2x})-\frac{5}{27} \phi_{1,x}^4 -\frac{2}{9} \phi_{1,x}^2
\phi_{2,x}^2-\frac{4}{9} \phi_{1,x}^3 \phi_{2,x}
-\frac{4}{27} \phi_{1,x} \phi_{2,x}^3\\
\qquad - e^{-2 \phi_2}- e^{-4 \phi_1-4 \phi_2}+ e^{2 \phi_1+2 \phi_2}+ e^{4
\phi_{2}}+\frac{1}{3}(6 \phi_{1,x}^2+12 \phi_{1,x}\phi_{2,x}+2 \phi_{1,2x}+4
\phi_{2,2x} ) e^{2 \phi_1}\\\qquad -\frac{1}{3}(2 \phi_{1,x}^2-4
\phi_{1,x}\phi_{2,x}-13 \phi_{2,x}^2-2 \phi_{1,2x}-10 \phi_{2,2x} ) e^{2
\phi_2}\\\qquad -\frac{1}{3}(7 \phi_{1,x}^2+22 \phi_{1,x}\phi_{2,x}+13 \phi_{2,x}^2-8
\phi_{1,2x}-10 \phi_{2,2x} ) e^{-2 \phi_1-2 \phi_2}
\end{array}
\end{eqnarray*}
and the second component $G(\phi_1,\phi_2)_{\phi_2}=-G(\phi_2,\phi_1)_{\phi_1}$.

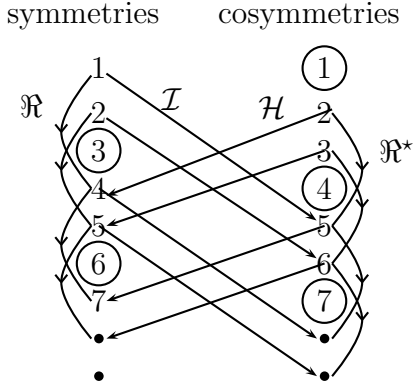
\begin{figure}
\begin{center}
\begin{pspicture}(0,0)(6,5)
\rput(0.8,4.8){symmetries} \rput(3.8,4.8){cosymmetries} \psdot*(1,0.0) \psdot*(1,0.5)
\rput(1,1){\rnode{G} 7} \cnodeput(1,1.5){F}{6} \rput(1,2.0){\rnode{E} 5}
\rput(1,2.5){\rnode{D} 4} \cnodeput(1,3.0){C}{3} \rput(1,3.5){\rnode{B} 2}
\rput(1,4.1){\rnode{A} 1} \psdot*(4,0.0) \psdot*(4,0.5) \cnodeput(4,1){N}{7}
\rput(4,1.5){\rnode{M}6} \rput(4,2.0){\rnode{L} 5} \cnodeput(4,2.5){K}{4}
\rput(4,3.0){\rnode{J} 3} \rput(4,3.5){\rnode{I} 2} \cnodeput(4,4.1){H}{1}
\ncline{->}{A}{L} \ncline{->}{B}{M}
\psline{->}(1,2.5)(3.9,0.5)\psline{->}(1,2)(3.9,0) \psline{->}(4.10,3.55)(1.1,2.4)
\psline{->}(4,3)(1.1,2.0)\psline{->}(4,2)(1.1,1.0)\psline{->}(4,1.5)(1.1,0.5)
\pscurve[](0.9,4)(0.5,3.25)(0.9,2.5)\psline(0.42,3.35)(0.5,3.25)
\psline(0.58,3.35)(0.5,3.25)
\pscurve[](0.9,3.5)(0.5,2.75)(0.9,2.0)\psline(0.42,2.85)(0.5,2.75)
\psline(0.58,2.85)(0.5,2.75)
\pscurve[](0.9,2.5)(0.5,1.75)(0.9,1.0)\psline(0.42,1.85)(0.5,1.75)
\psline(0.58,1.85)(0.5,1.75)
\pscurve[](0.9,2.0)(0.5,1.25)(0.9,0.5)\psline(0.42,1.35)(0.5,1.25)
\psline(0.58,1.35)(0.5,1.25)
\pscurve[](4.1,3.5)(4.5,2.75)(4.1,2.0)\psline(4.42,2.85)(4.5,2.75)
\psline(4.58,2.85)(4.5,2.75)
\pscurve[](4.1,3.0)(4.5,2.25)(4.1,1.5)\psline(4.42,2.35)(4.5,2.25)
\psline(4.58,2.35)(4.5,2.25)
\pscurve[](4.1,2.0)(4.5,1.25)(4.1,0.5)\psline(4.42,1.35)(4.5,1.25)
\psline(4.58,1.35)(4.5,1.25)
\pscurve[](4.1,1.5)(4.5,0.75)(4.1,0.0)\psline(4.42,0.85)(4.5,0.75)
\psline(4.58,0.85)(4.5,0.75) \rput(1.97,3.68){$\J$}\rput(3.3,3.58){$\cH$}
\rput(0.10,3.6){$\Re$}\rput(4.96,3.0){$\Re^{\star}$}
\end{pspicture}
\end{center}
\caption{Interrelations between $\Re$ and $\cH, \J$ of system
(\ref{sys})}\label{fig1}
\end{figure}

The adjoint operator of $\Re^{\star}$ gives rise to the cosymmetries of order $3n+2$
and $3n+3$. In figure \ref{fig1}, we list out the orders of symmetries in the left
row and orders of cosymmetries in the right row. We use a circle around a number $k$
to denote that the system does not possess the symmetries or cosymmetries of order
$k$.

\section{Symplectic and Hamiltonian structures}\label{sec5}
In this section, we show system (\ref{sys}) is generalised bi-Hamiltonian by
presenting its Hamiltonian and symplectic operators. Surprisingly, the product of
these two operator does not lead to the recursion operator we constructed in section
\ref{sec42}, but to its square.

We know Hamiltonian operators map cosymmetries to symmetries while symplectic
operators map symmetries to cosymmetries. For system (\ref{sys}), from Figure
\ref{fig1} we can draw the conclusion that the possible order of Hamiltonian operators
can only be $3k+2$ and of symplectic operators $3k+1$. Here we
consider positive orders, i.e., $k\geq 0$.

For system (\ref{sys}), there exists an anti-symmetric operator $\cH$ such that
\begin{eqnarray}\label{ha1}
\left(\begin{array}{lr} \cH_{11} & \cH_{12}\\\cH_{21} & \cH_{22} \end{array} \right)
\left(\begin{array}{l} s_{1,\phi_1}\\s_{1,\phi_2} \end{array} \right) =-9
\left(\begin{array}{lr} \Re_{11} & \Re_{12}\\\Re_{21} & \Re_{22} \end{array} \right)
\left(\begin{array}{l} \phi_{1,x}\\\phi_{2,x} \end{array} \right),
\end{eqnarray}
where $\Re$, $s_1^1$ and $s_1^2$ are defined in Theorem \ref{ReS} and
\begin{eqnarray*}
\begin{array}{ll}
{\cal H}_{11}=\phi_{1,x} D_x+D_x \phi_{1,x} +2 \phi_{2,x} D_x +2 D_x \phi_{2,x}
-\phi_{1,x} D_x^{-1} \phi_{1,t} -\phi_{1,t} D_x^{-1} \phi_{1,x}\\
{\cal H}_{12}=3 D_x^2 +2 \phi_{1,x} D_x -2 \phi_{2,x} D_x +\phi_{1,xx} -\phi_{2,xx}
-\phi_{1,x}^2-\phi_{2,x}^2 - \phi_{1,x} \phi_{2,x} \\
\qquad\ +3 (v_1^2+v_2^2+v_3^2) -\phi_{1,x} D_x^{-1} \phi_{2,t} -\phi_{1,t} D_x^{-1}
\phi_{2,x}
\end{array}
\end{eqnarray*}
and ${\cal H}_{21}(\phi_1,\phi_2)=-{\cal H}_{12}(\phi_2,\phi_1)$; ${\cal H}_{22}
(\phi_1, \phi_2)=-{\cal H}_{11}(\phi_2, \phi_1)$.
\begin{The}\label{ham}
The operator $\cH$ defined above is Hamiltonian.
\end{The}
{\bf Proof}. We prove this by checking the condition of Theorem 7.8 in
\cite{mr94g:58260}. The associated bi-vector of $\cH$ is by definition
$$\Theta= \frac{1}{2} \int \theta \wedge \cH \theta,\quad \mbox{where}\quad
\theta=\left(\begin{array}{c} \theta_1, \theta_2\end{array}\right)^T.$$ We need to
check the vanishing of the tri-vector: $\Pr {\bf v}_{\cH \theta} (\Theta)=0.$ Instead of
writing out the full calculation, we pick out terms with highest degree in
$x$-derivatives of dependant variables $\phi_1$ and $\phi_2$ in 3-form of $\theta_1$.
The relevant terms in $\cH \theta$ are
\begin{eqnarray*}
\left(\begin{array}{c}-\phi_{1,x} \left((2 \phi_{1,x} \phi_{2,x}
+\phi_{1,x}^2)\theta_1\right)_{-1} -(2 \phi_{1,x} \phi_{2,x} +\phi_{1,x}^2)
\left(\phi_{1,x}\theta_1\right)_{-1}\\
-\phi_{2,x} \left( (2 \phi_{1,x} \phi_{2,x} +\phi_{1,x}^2)\theta_1\right)_{-1} + (2
\phi_{1,x} \phi_{2,x} +\phi_{2,x}^2) \left(\phi_{1,x}\theta_1\right)_{-1}
\end{array}\right),
\end{eqnarray*}
where $(p)_{-1}$ denotes $D_x^{-1}(p)$ and such terms in $\Theta$ are
\begin{eqnarray*}
\begin{array}{l}
\quad \int -\phi_{1,x} \theta_1 \wedge \left((2 \phi_{1,x} \phi_{2,x}
+\phi_{1,x}^2)\theta_1\right)_{-1} =\int \phi_{1} \theta_{1,x} \wedge \left((2
\phi_{1,x} \phi_{2,x} +\phi_{1,x}^2)\theta_1\right)_{-1}\\ =-\int (2 \phi_{1,x}
\phi_{2,x} +\phi_{1,x}^2)\theta_1 \wedge \left(\phi_{1,x}\theta_1\right)_{-1}
\end{array}
\end{eqnarray*}
using integration by parts. Thus the terms we look at in $\Pr {\bf v}_{\cH \theta}
(\Theta)$ are
\begin{eqnarray*}
\begin{array}{l}
\quad \int -(2 \phi_{1,x} \phi_{2,x} +\phi_{1,x}^2)
\left(\phi_{1,x}\theta_1\right)_{-1}\wedge \theta_{1,x} \wedge \left((2 \phi_{1,x}
\phi_{2,x} +\phi_{1,x}^2)\theta_1\right)_{-1}\\ +  \int 2 (\phi_{2,x} +\phi_{1,x} )
\phi_{1,xx} \left((2 \phi_{1,x} \phi_{2,x} +\phi_{1,x}^2)\theta_1\right)_{-1}  \wedge
\theta_1 \wedge \left(\phi_{1,x}\theta_1\right)_{-1}\\+\int  2 \phi_{1,x} \phi_{2,xx}
\left( (2 \phi_{1,x} \phi_{2,x} +\phi_{1,x}^2)\theta_1\right)_{-1}  \wedge \theta_1
\wedge \left(\phi_{1,x}\theta_1\right)_{-1}\\
=0.
\end{array}
\end{eqnarray*}
Similarly we can prove that the tri-vector $\Pr {\bf v}_{\cH \theta} (\Theta)$ vanishes,
which implies that $\cH$ is a Hamiltonian operator. \hfill $\blacksquare$

We now construct a symplectic operator of system (\ref{sys}). Its lowest positive order
is $1$.
\begin{Pro} No weakly nonlocal symplectic operator of order $1$ exists for system
(\ref{sys}).

\end{Pro}
{\bf Proof}. Assume that any of the nonlocal terms of the symplectic operator is of
the form  $ \xi_1 D_x^{-1}\xi_2 $, where $\xi_i\in \Omega^1$. Since system
(\ref{sys}) is homogeneous in the variables $\phi_i$ and $v_i$, $\xi_i$ is also
homogeneous; its possible weights are $0, 1$. From the results in \cite{mr1974732},
the $\xi_i$ are cosymmetries of the system. One can check that the system has no
cosymmetries of weight $1$ and $0$, that is, no conserved densities of weight $0$ and $1$.
\hfill $\blacksquare$

We now look for next lowest order symplectic operator, which is $4$. Indeed, there
exists an operator $\J$ such that
\begin{eqnarray}\label{sy1}
\left(\begin{array}{lr} \J_{11} & \J_{12}\\\J_{21} & \J_{22} \end{array} \right)
\left(\begin{array}{l} \phi_{1,t}\\\phi_{2,t} \end{array} \right) =\delta \!\int\!\!
g,
\end{eqnarray}
where \begin{eqnarray*}
\begin{array}{l}
g=3 \phi_{1,3x}^2+3 \phi_{1,3x} \phi_{2,3x}+3 \phi_{2,3x}^2 -\frac{15}{2} \phi_{1,xx}
\phi_{2,xx} (\phi_{1,xx}+\phi_{2,xx})-(24 v_1^2+15 v_2^2+24 v_3^2)
\phi_{1,xx}^2\\\qquad+5 (\phi_{1,x}^2+\phi_{1,x}
\phi_{2,x}+\phi_{2,x}^2)(\phi_{1,xx}^2+\phi_{1,xx} \phi_{2,xx}+\phi_{2,xx}^2)-(15
v_1^2+24 v_2^2+24 v_3^2) \phi_{2,xx}^2\\\qquad-(15 v_1^2+15 v_2^2+33 v_3^2)
\phi_{1,xx} \phi_{2,xx}+\frac{15}{2} (2v_1^2-v_2^2-v_3^2) \phi_{1,xx} \phi_{2,x}^2
\\\qquad+\frac{15}{2} (2v_2^2-v_1^2-v_3^2) \phi_{1,x}^2 \phi_{2,xx}-\frac{5}{2} \phi_{1,x}^2
\phi_{2,x}^2 (\phi_{1,xx}+\phi_{2,xx})+\frac{5}{6} (\phi_{1,x}^4 \phi_{2,xx} +
\phi_{1,xx}\phi_{2,x}^4)\\\qquad+\frac{7}{27} (\phi_{1,x}^6+\phi_{2,x}^6)
+\frac{7}{9}(\phi_{1,x}^5 \phi_{2,x}+\phi_{1,x} \phi_{2,x}^5) +\frac{5}{9}( 3 v_2^2+
\phi_{2,x}^2)\phi_{1,x}^4+\frac{5}{9}(3 v_1^2+ \phi_{1,x}^2)\phi_{2,x}^4
\\\qquad+\frac{89}{3} (v_1^2+v_3^2) \phi_{1,x}^4+ \frac{89}{3} (v_2^2+v_3^2)
\phi_{2,x}^4-\frac{5}{27} \phi_{1,x}^3 \phi_{2,x}^3 +(\frac{421}{3}
v_3^2-\frac{65}{3} v_2^2+\frac{10}{3} v_1^2) \phi_{1,x} \phi_{2,x}^3\\\qquad
+(\frac{421}{3} v_3^2-\frac{65}{3} v_1^2+\frac{10}{3} v_2^2) \phi_{2,x}
\phi_{1,x}^3+(223v_3^2-5 v_1^2 -5 v_2^2) \phi_{1,x}^2 \phi_{2,x}^2\\\qquad  +(15
v_2^4 +60 v_1^4 +60 v_3^4 +66 v_1^2 v_2^2 +66 v_2^2 v_3^2 +84 v_1^2 v_3^2)
\phi_{1,x}^2\\\qquad+(15 v_1^4 +60 v_2^4 +60 v_3^4 +66 v_1^2 v_2^2 +66 v_1^2 v_3^2
+84 v_2^2 v_3^2) \phi_{2,x}^2\\\qquad+ (15 v_1^4 +15 v_2^4 +105 v_3^4 +48 v_1^2 v_2^2
+84 v_1^2 v_3^2 +84 v_2^2 v_3^2) \phi_{1,x} \phi_{2,x}-3(v_1^2+v_2^2+v_3^2)^3
\end{array}
\end{eqnarray*}
and
\begin{eqnarray*}
\begin{array}{ll}
{\J}_{11}= (2 \phi_{1,x} + \phi_{2,x}) D_x^3 +D_x^3 (2 \phi_{1,x} + \phi_{2,x})+q D_x
+D_x q +\frac{1}{3} ( s_{1,\phi_1} D_x^{-1} s_{2,\phi_1}+ s_{2,\phi_1}
D_x^{-1} s_{1,\phi_1});\\
{\J}_{12}=3 D_x^4 +2 ( \phi_{1,x} -\phi_{2,x}) D_x^3 + p_2 D_x^2 + p_1 D_x + p_0 +
\frac{1}{3} ( s_{1,\phi_1} D_x^{-1} s_{2,\phi_2}+ s_{2,\phi_1} D_x^{-1}
s_{1,\phi_2}).
\end{array}
\end{eqnarray*}
Here $s_{j,\phi_1}$, $s_{j,\phi_2}$ are the components of cosymmetries and
\begin{equation*}
\begin{array}{l}
q=-2 \phi_{1,xxx}-\phi_{2,xxx}- 2 \phi_{1,x} \phi_{2,xx}- \phi_{2,x} \phi_{2,xx}
-\frac{4}{3} \phi_{1,x}^3 -2 \phi_{1,x}^2 \phi_{2,x} +\frac{1}{3} \phi_{2,x}^3\\
\qquad+ (13\phi_{2,x}-4 \phi_{1,x}) v_1^2 - (4 \phi_{1,x} + 2 \phi_{2,x}) v_2^2-(4
\phi_{1,x}+17 \phi_{2,x}) v_3^2 ;
\\ p_2=3 \phi_{1,xx}-3 \phi_{2,xx}-\phi_{1,x}^2 -\phi_{1,x} \phi_{2,x} -\phi_{2,x}^2
+15 (v_1^2 + v_2^2  +v_3^2);
\\ p_1=\phi_{1,xxx}-\phi_{2,xxx}- 2 \phi_{1,x} \phi_{2,xx}-4\phi_{2,x} \phi_{2,xx}
-\frac{4}{3} \phi_{1,x}^3 - \phi_{1,x}^2 \phi_{2,x} +\phi_{1,x} \phi_{2,x}^2
+\frac{4}{3} \phi_{2,x}^3\\
\qquad+ (11\phi_{1,x}+4 \phi_{2,x}) v_1^2 - (4 \phi_{1,x} -49 \phi_{2,x}) v_2^2-(19
\phi_{1,x}+41 \phi_{2,x}) v_3^2 ;
\\ p_0=-(\phi_{1,x}+2 \phi_{2,x}) \phi_{2,xxx}+ 2 \phi_{1,xx}^2+\phi_{1,xx} \phi_{2,xx}
- 2 (\phi_{1,x}^2 +\phi_{1,x} \phi_{2,x}) \phi_{1,xx}  -
(\phi_{1,x}^2 - 2 \phi_{2,x}^2 )\phi_{2,xx}\\
\qquad+ (10 \phi_{1,xx}+2 \phi_{2,xx}-2 \phi_{1,x}^2 - 4 \phi_{2,x}^2 ) v_1^2 - (2
\phi_{1,xx}-29 \phi_{2,xx}+4 \phi_{1,x}^2 +8\phi_{1,x} \phi_{2,x} -36
\phi_{2,x}^2) v_2^2\\
\qquad-(14 \phi_{1,xx}+25 \phi_{2,xx}- 6 \phi_{1,x}^2 - 38 \phi_{1,x} \phi_{2,x} - 28
\phi_{2,x}^2) v_3^2 + 12 (v_1^2 + v_2^2  + v_3^2)^2;
\end{array}
\end{equation*}
And ${\J}_{21}(\phi_1,\phi_2)=-{\J}_{12}(\phi_2,\phi_1)$;
${\J}_{22}(\phi_1,\phi_2)=-{\J}_{11}(\phi_2,\phi_1)$.
\begin{The}\label{symp}
The operator $\J$ defined above is symplectic.
\end{The}
{\bf Proof}. We use the notation $\d \phi=(\d \phi_1,\ \d \phi_2)^T$. The $2$-form
defined by the operator $\J$ is
\begin{eqnarray*}
\begin{array}{l}
\omega=\frac{1}{2} \int \d \phi \wedge \J \d \phi \\
\quad = \int \left((2 \phi_{1,x} + \phi_{2,x}) \d \phi_1 \wedge \d
\phi_{1,3x}+q(\phi_1,\phi_2) \ \d \phi_1 \wedge \d \phi_{1,x} +\frac{1}{3}
s_{1,\phi_{1}} \d \phi_1 \wedge D_x^{-1} (s_{2,\phi_{1}} \d \phi_1)\right.\\
\qquad +3\ \d \phi_1 \wedge \d \phi_{2,4x} +2 ( \phi_{1,x} -\phi_{2,x}) \d \phi_1
\wedge \d \phi_{2,3x} + p_2 \d \phi_1 \wedge \d \phi_{2,2x} + p_1 \d \phi_1 \wedge \d
\phi_{2,x} \\
\qquad + p_0 \d \phi_1 \wedge \d \phi_{2}+ \frac{1}{3}  s_{1,\phi_{1}}\d \phi_{1}
D_x^{-1} (s_{2,\phi_{2}} \d \phi_{2} )+ \frac{1}{3} s_{2,\phi_{1}} \d \phi_{1}
D_x^{-1} (s_{1,\phi_{2}} \d \phi_{2} )\\
\qquad\left.-(2 \phi_{2,x} + \phi_{1,x}) \d \phi_2 \wedge \d
\phi_{2,3x}-q(\phi_2,\phi_1) \ \d \phi_2 \wedge \d \phi_{2,x} +\frac{1}{3}
s_{1,\phi_{2}} \d \phi_2 \wedge D_x^{-1} (s_{2,\phi_{2}} \d \phi_2) \right)\ .
\end{array}
\end{eqnarray*}
We now compute $\d \omega$. Instead of carrying out the whole computation, we only
demonstrate the method by picking out $3$-forms of $\phi_{1}$ and its $x$-derivatives
in $\d \omega$. These are as follows:
\begin{eqnarray*}
\begin{array}{l}
\qquad\int \left(2 \d \phi_{1,x}\wedge \d \phi_1 \wedge \d \phi_{1,3x} -2 \d
\phi_{1,3x} \wedge d \phi_1 \wedge d \phi_{1,x}-\frac{2}{3} \d \phi_{1,xx}\wedge \d
\phi_1 \wedge D_x^{-1}
(s_{2,\phi_{1}} \d \phi_1)\right.\\
\quad\left.-\frac{1}{3} s_{1,\phi_{1}} \d \phi_1 \wedge D_x^{-1} ((4\phi_{1,x}+2
\phi_{2,x}) \d \phi_{1,xx} \wedge \d \phi_1+(4 \phi_{1,xx}+2 \phi_{2,xx}) \d
\phi_{1,x}\wedge \d \phi_1) \right)\\
\quad=\int \left(\frac{2}{3} s_{2,\phi_{1}} D_x^{-1} (\d \phi_{1,xx}\wedge \d \phi_1)
\wedge \d \phi_1 -\frac{1}{3} s_{1,\phi_{1}} \d \phi_1 \wedge (4\phi_{1,x}+2
\phi_{2,x}) \d \phi_{1,x} \wedge \d \phi_1\right)\\
\quad=\int \frac{2}{3} s_{2,\phi_{1}}  \d \phi_{1,x}\wedge \d \phi_1 \wedge \d \phi_1
=0.
\end{array}
\end{eqnarray*}
By working out for other terms, we can show $\d \omega=0$. Thus we prove $\J$ is
symplectic. $\blacksquare$

\begin{The}System (\ref{sys}) is a bi-Hamiltonian system.
\end{The}
{\bf Proof}. We only need to show that the symplectic operator $\J$ defined in
Theorem \ref{symp} and Hamiltonian operator $\cH$ defined in Theorem \ref{ham} are
compatible. Operator $\cH$ is of order $2$ and operator $\J$ is of order $4$. This
leads to that $\cH \J$ is of order $6$. By straightforward computation, one can
verifies that $\cH \J = 27 \Re^2$, where $\Re$ is defined in Theorem \ref{ReS}. From
Theorem \ref{RN}, we know $\Re$ is Nijenhuis. So is the operator $\Re^2$. Thus these
two operators are compatible. Hence we proved the statement. \hfill $\blacksquare$

We can check that the conditions in Theorem \ref{th1} are satisfied for $\cH$ and
$\J$. Therefore, starting from four starting points, two symmetries appeared in $\cH$
and two cosymmetries appeared in $\J$, we can generate a hierarchy of commuting local
symmetries, which are all Hamiltonian vector fields and their Hamiltonian are in
involution.

\section{Discussion}\label{sec6}
In this paper, we prove that for compatible weakly nonlocal Hamiltonian and
symplectic operators, hierarchies of infinitely many commuting local symmetries and
conservation laws can be generated under some easily verified conditions no matter
whether the generating Nijenhuis operators are weakly nonlocal or not. The problem
how to generate local symmetries and conservation laws when the Nijenhuis operators
are no longer weakly nonlocal has not been studied before. As in Example \ref{eg1}
and \ref{eg2} where the objects are differential polynomials, we believe that
assumption \ref{con4} in Theorem \ref{th2} can be relaxed in general. However, we are
not able to simplify this assumption yet.

We construct a recursion operator $\Re$, a Hamiltonian operator $\cH$ and a
symplectic operator $\J$ for system (\ref{sys}). We show that $\cH \J =\Re^2$. This
leads to $\Re^{2 k} \cH$ being Hamiltonian and compatible to $\cH$ for $k\in
\mathbb{N}$. An immediate question is: is $\Re \cH$ Hamiltonian and
compatible to $H$? We conjecture the answer is positive. However, the computation
involved is rather big and we have not found an elegant way to prove it.

In the Lax representation $L(\la)$ of system (\ref{sys}), cf. formula (\ref{lax}),
$\dl$ is a $3\times 3$ matrix. The natural generalisation is $\dl$ being $n\times n$
matrix. The construction in this paper works for any finite $n$. For a given $n$, the
corresponding system possesses a recursion operator $\Re$ of order $n$, which can be
constructed in the same manner as in Section \ref{sec42}. The system is
bi-Hamiltonian with the lowest positive order of Hamiltonian operator $\cH$ being
$n-1$ and that of a symplectic operator $\J$ being $n+1$. These operators have the
same properties as we discussed for $n=3$, namely, $\cH \J$ does not give rise to
$\Re$, but $\Re^2$.

If we treat arbitrary $n$ by considering $\dl$ as a shift operator, this leads to a
$2+1$-dimensional lattice-field integrable equation. Recently, Blaszak and Szum have
constructed Hamiltonian operators for such type of equations with a certain type of
Lax operator \cite{BS01}.  It would be interesting to construct the bi-Hamiltonian
structure for the $2+1$-dimensional lattice-field equation and to see how this
structure is related to the ones with finite periods.

\section*{Acknowledgements}
The author would like to thank A.N.W. Hone, A.V. Mikhailov, V.S. Novikov, P.J. Olver,
J.A. Sanders and A. Sergyeyev for inspiring discussions and useful comments.

\section*{Appendix}
Here we give the lemmas used in proof of Theorem \ref{th1} and Theorem \ref{th2}.
Lemma \ref{lem4} is due to Sergyeyev \cite{serg5}. To increase the readability, we
include the statement here.
\begin{Lem}\label{lem4}
Let the nonlocal terms of a Nijenhuis operator $\Re:\lieh \rightarrow \lieh$ be of
the form
$$\begin{array}{l}\sum_{j=1}^l Q_j \otimes D_x^{-1} \beta_j, \quad \mbox{where} \ \
P_j\in \lieh\ \ \mbox{and} \ \ \beta_j\in \Omega^1. \end{array}$$
\begin{enumerate}
\item If for all $j,k=1,\cdots , l$, we have $L_{Q_j}\beta_k=0$ and both $
\beta_j$ and $\Re^{\star}(\beta_j)$ are closed, then for any $h_0\in \lieh$ such that
$L_{h_0} \Re=0$ and $L_{h_0} \beta_j=0$, all $h_i=\Re^i (h_0)$ are local and commute,
where $i=0,1, 2, \cdots $. \label{flow}
\item If $L_{Q_j}\Re=0$, then for any $\xi_0$ such that $L_{Q_j}\xi_0=0$ and both $
\xi_0$ and $\Re^{\star}(\xi_0)$ are closed,  all $\xi_i=\Re^{\star i} (\xi_0)$ are
local and closed, where $i=0,1, 2, \cdots $. \label{form}
\end{enumerate}
\end{Lem}

The following theorem was proved by A.V. Mikhailov, Applied Mathematics Department,
University of Leeds and published here with his kind permission.
\begin{The}\label{apth}
Let $\f_k,\g_k \in\A^N$ (either in $\lieh$ and/or in $\Omega^1$), vector-columns
$\f_1,\, \ldots\, ,\f_n$ be linearly independent over $\bbbc$ and
\begin{equation}
 \label{1}
\sum_{k=1}^n \f_k D_x^{-1}\g_k^{\tr}+\g_k D_x^{-1}\f_k^{\tr}=0,
\end{equation}
then
\[ \g_k=\sum_{i=1}^n \f_i A_{i\, k}\]
where
\[ A _{i\, k}=-A _{k\, i},\quad D_x(A _{i\, k})=0.\]
\end{The}
The proof of the Theorem is based on two Lemmas.
\begin{Lem}
 If
\[ \sum_{k=1}^n \f_k D_x^{-1}\g_k^{\tr}+\g_k D_x^{-1}\f_k^{\tr}=0\]
then for any $p,q\in\bbbz_{\ge 0}$
\begin{equation}
 \label{2} \sum_{k=1}^n D_x^p(\f_k) D_x^{q}(\g_k^{\tr})+D_x^p(\g_k) D_x^{q}(\f_k^{\tr})=0.
\end{equation}
\end{Lem}

{\bf Proof}. Indeed, it follows from (\ref{1}) that
\[
 \sum_{j=0}^\infty (-1)^j \left(\sum_{k=1}^n \f_k D_x^{j}\g_k^{\tr}+\g_k D_x^{j}
 \f_k^{\tr}\right)D_x^{-1-j}=0
\]
and thus for any $q\in\bbbz_{\ge 0}$
\begin{equation}
 \label{3}\sum_{k=1}^n \f_k D_x^{q}(\g_k^{\tr})+\g_k D_x^{q}(\f_k^{\tr})=0,
 \quad q=0,1,\ldots\, .
\end{equation}
In order to demonstrate (\ref{2}) we use induction. It follows from (\ref{3}) that
the statement is true for $p=0$ and any $q\in \bbbz_{\ge 0}$. Let us assume that
(\ref{2}) is valid for any $q$ and $p\le l-1$ and then show that it is true for
$p=l$. For $p= l-1$ we have
\[ \sum_{k=1}^n D_x^{l-1}(\f_k) D_x^{q}(\g_k^{\tr})+D_x^{l-1}(\g_k) D_x^{q}(\f_k^{\tr})=0.
\]
Applying $D_x$ we get
\[
 \sum_{k=1}^n\left( D_x^{l}(\f_k) D_x^{q}(\g_k^{\tr})+D_x^{l}(\g_k)
 D_x^{q}(\f_k^{\tr})\right)+
\sum_{k=1}^n \left(D_x^{l-1}(\f_k) D_x^{q+1}(\g_k^{\tr})+D_x^{l-1}(\g_k)
D_x^{q+1}(\f_k^{\tr})\right)=0.
\]
The last sum vanishes due to the induction assumption. Thus (\ref{2}) is true for
$p=l$.\hfill $\blacksquare$

Let ${\bf F}_k$, ${\bf G}_k$ denote infinite dimensional vector-columns
\[ {\bf F}_k=\left(\begin{array}{c}
 \f_k\\
D_x(\f_k)\\
\vdots \\
D_x^p(\f_k)\\
\vdots
\end{array}\right)\, ,\qquad
{\bf G}_k=\left(\begin{array}{c}
 \g_k\\
D_x(\g_k)\\
\vdots \\
D_x^p(\g_k)\\
\vdots
\end{array}\right)\, ,
\]
and $F,G$ denote matrices
\[ F=({\bf F}_1,\, \ldots\, , {\bf F}_n),\qquad G=({\bf G}_1,\, \ldots\, , {\bf G}_n).\]
Then (\ref{2}) can be written in the form
\begin{equation}
 \label{4}
FG^{\tr}+GF^{\tr}=0\, .
\end{equation}

\begin{Lem}
 Let vectors $\f_1,\ldots , \f_n$ be linearly independent over $\bbbc$,
 then vectors ${\bf F}_1,\ldots ,{\bf F}_n$ are linearly independent over $\A$ and thus
\[ \mbox{\rm\rank} (F)=n.\]
\end{Lem}

{\bf Proof}. Let us assume the opposite, i.e. $\rank (F)=m<n$. Without a loss of
generality we shall assume that the first $m$ vectors ${\bf F}_1,\ldots ,{\bf F}_m$
are linearly independent over $\A$ and thus the rest vectors ${\bf F}_k, \
k=m+1,\ldots n$ can be expressed as
\[ {\bf F}_k=\sum_{s=1}^m {\bf F}_s\alpha_{sk},\quad \alpha_{sk}\in\A,
\quad k=m+1,\ldots ,n.\] The latter is equivalent to
\begin{equation}
 \label{5}
D_x^p(\f_k)=\sum_{s=1}^m D_x^p(\f_s)\alpha_{sk},\quad k=m+1,\ldots ,n, \quad
p\in\bbbz_{\ge 0}.\end{equation} Differentiating (\ref{5}) we find
\[  D_x^{p+1}(\f_k)=\sum_{s=1}^m D_x^p(\f_s)D_x(\alpha_{sk})
+\sum_{s=1}^m D_x^{p+1}(\f_s)\alpha_{sk}=
 \sum_{s=1}^m D_x^p\f_sD_x(\alpha_{sk})+D_x^{p+1}(\f_k)
\]
and thus
\[ \sum_{s=1}^m D_x^p(\f_s)D_x(\alpha_{sk})=0.\]
Since vectors ${\bf F}_1,\ldots ,{\bf F}_m$ are assumed to be linearly independent,
we have $D_x(\alpha_{sk})=0$ and thus $\alpha_{sk}\in\bbbc$.  It follows from
(\ref{5}) at $p=0$ that vectors $\f_k$ are linearly dependent over $\bbbc$.\hfill
$\blacksquare$

{\bf Proof of Theorem \ref{apth}}. From Lemma 2 it follows that the rank of matrix
$F$ in (\ref{4}) is $n$, thus vector columns of $F$ and $G$ span the same linear
space and therefore
\begin{equation}
\label{6} {\bf G}_k=\sum_{i=1}^n {\bf F}_i A_{i\, k}
\end{equation}
or
\begin{equation}
\label{7} D_x^p(\g_k)=\sum_{i=1}^n D_x^p({\f}_i) A_{i\, k}\end{equation} Substitution
of (\ref{6}) in (\ref{4}) gives $F(A^{\tr}+A)F^{\tr}=0$ and since $\rank F=n$ we get
$A^{\tr}+A=0$. From (\ref{7}) it follows that
\[
 D_x^{p+1}(\g_k)=\sum_{i=1}^n D_x^p({\f}_i) D_x(A_{i\, k})
 +D_x^{p+1}({\f}_i) A_{i\, k}=D_x^{p+1}(\g_k)+\sum_{i=1}^n D_x^p({\f}_i) D_x(A_{i\, k})
\]
which leads to $D_x(A_{i\, k})=0$ (since $\rank F=n$). \hfill $\blacksquare$


\begin{thebibliography}{10}

\bibitem{AKNS74}
M.J. Ablowitz, D.J. Kaup, A.C. Newell and H.~Segur, Stud. Appl. Math. {\bf 53}, 249
(1974).

\bibitem{Olv77}
P.J. Olver, J. Math. Phys. {\bf 18}, 1212 (1977).

\bibitem{mr94h:35241}
C.~Athorne and I.~Ya. Dorfman, J. Math. Phys. {\bf 34}, 3507 (1993).

\bibitem{Fuc79}
B.~Fuchssteiner, Nonlinear Anal. Theor. Meth. Appl. {\bf 3}, 849 (1979).

\bibitem{Mag80}
F.~Magri, {\em Lecture Notes in Physics} {\bf 120}, (Springer, 1980), p. 233.

\bibitem{Mag78}
F.~Magri, J. Math. Phys. {\bf 19}, 1156 (1978).

\bibitem{GD79}
I.M. Gel'fand and I.Ya. Dorfman, Funct. Anal. Appl. {\bf 13}, 248 (1979).

\bibitem{mr82g:58039}
A.~S. Fokas and B.~Fuchssteiner, Lett. Nuovo Cimento (2) {\bf 28}, 299 (1980).

\bibitem{mr84j:58046}
B.~Fuchssteiner and A.~S. Fokas, Physica D {\bf 4}, 47 (1981).

\bibitem{ps05}
J.~Praught and G.~Smirnov, SIGMA {\bf 1}, 005 (2005).

\bibitem{MaN01}
A.Ya. Maltsev and S.P. Novikov, Physica D {\bf156}, 53 (2001).

\bibitem{wang1}
J.P. Wang, J. Nonlinear Math. Phys. {\bf 9},  213 (2002).

\bibitem{KS4}
A.~Karasu (Kalkanli), A.~Karasu and S.Y. Sakovich, Acta Appl. Math. {\bf 83}, 85
(2004).

\bibitem{serg5b}
A.~Sergyeyev, J. Phys. A {\bf 38},  L257 (2005).

\bibitem{mr94j:58081}
I. Dorfman, {\em Dirac structures and integrability of nonlinear evolution
  equations} (John Wiley \& Sons Ltd., Chichester, 1993).

\bibitem{dorf87}
I.Y. Dorfman, Phys. Lett. A {\bf 125}, 240 (1987).

\bibitem{mr2001h:37146}
M. G{\"u}rses, A. Karasu and V.~V. Sokolov, J. Math. Phys. {\bf 40}, 6473 (1999).

\bibitem{mik79}
A.V. Mikhailov, JETP Lett. {\bf 30}, 414 (1979).

\bibitem{mik81}
A.V. Mikhailov, Physica D {\bf 3}, 73 (1981).

\bibitem{LM04}
S.~Lombardo and A.V. Mikhailov, J. Phys. A {\bf 37}, 7727 (2004).

\bibitem{LM05}
S.~Lombardo and A.V. Mikhailov, Comm. Math. Phys. {\bf 258}, 179  (2005).

\bibitem{mr90b:58085}
P.~J. Olver and Y. Nutku, J. Math. Phys. {\bf 29},  1610 (1988).

\bibitem{mr89g:58092}
A.~V. Mikhailov, A.~B. Shabat and R.~I. Yamilov, Comm. Math. Phys. {\bf 115}, 1
(1988).

\bibitem{FS88c}
P.~M. Santini and A.~S. Fokas, Commun. Math. Phys. {\bf 115}, 375 (1988).

\bibitem{FS88b}
A.~S. Fokas and P.~M. Santini, Commun. Math. Phys. {\bf 116}, 449 (1988).

\bibitem{bm}
R.~Bury and A.V. Mikhailov, {\em Solitons and wave fronts in periodic two dimensional
volterra system.}, In preparation (2008).

\bibitem{wang98}
J.~P. Wang, {\em Symmetries and Conservation Laws of Evolution Equations}, PhD
thesis, Amsterdam, 1998.


\bibitem{mr94g:58260}
P.~J. Olver, {\em Applications of {L}ie groups to differential equations}, 2nd ed.
(Springer-Verlag, New York, 1993).

\bibitem{Mal2}
A.Ya. Maltsev, Int. J. Math. and Math. Sci. {\bf 30}, 393 (2002).

\bibitem{mr1974732}
J.~A. Sanders and J. P. Wang, Nonlinear Anal. Theor. Meth. Appl. {\bf 47}, 5213
(2001).

\bibitem{serg5}
A.~Sergyeyev, J. Phys. A  {\bf 38}, 3397 (2005).

\bibitem{mnw07}
A.V. Mikhailov, V.S. Novikov and J.~P. Wang, Stud. Appl. Math. {\bf 118},  419
(2007).

\bibitem{mr83d:58031}
B. Fuchssteiner and W. Oevel, J. Math. Phys. {\bf 23}, 358 (1982).

\bibitem{serg7b}
A.~Sergyeyev and D.~Demskoi, J. Math. Phys. {\bf 48}, 042702 (2007).

\bibitem{mr95k:35154}
S.~I. Svinolupov and V.~V. Sokolov, Teoret. Mat. Fiz. {\bf 100}, 214 (1994).

\bibitem{anco6}
S.C. Anco, SIGMA {\bf 2}, 044 (2006).

\bibitem{mr88g:58080}
B. Fuchssteiner, W. Oevel and W. Wiwianka, Comput. Phys. Comm. {\bf 44},  47 (1987).

\bibitem{gu98}
{\"U}.~G{\"o}kta\c{s}, {\em Algorithmic Computation of Symmetries, Invariants and
Recursion Operators for Systems of Nonlinear Evolution and Differential-Difference
  Equations}, PhD thesis, Golden, 1998.

\bibitem{mr2002b:37100}
J.~A. Sanders and J.~P. Wang, Physica D {\bf 149}, 1 (2001).

\bibitem{BS01}
 M.~Blaszak and A.~Szum, J. Math. Phys. {\bf 42}, 225 (2001).


\end{thebibliography}
\end{document}